\documentclass[fleqn,usenatbib]{mnras}

\usepackage{newtxtext,newtxmath}
\usepackage[T1]{fontenc}

\usepackage{graphicx}	% Including figure files
\usepackage{amsmath}	% Advanced maths commands
\usepackage{placeins}
\usepackage{subcaption}
\usepackage{caption}
\usepackage{xcolor}
\newcommand{\orcid}[1]{\textsuperscript{\,\,\href{https://orcid.org/#1}{\includegraphics[scale=0.06]{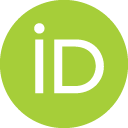}}}}

\newcommand{\HA}{\mbox{H$\alpha$}}
\newcommand{\NII}{[\mbox{N{\sc{ii}}]}}
\newcommand{\HII}{\mbox{H{\sc{ii}}}}

\newcommand{\NH}{[\mbox{N{\sc{ii}}]/H$\alpha$}}
\newcommand{\SH}{[\mbox{S{\sc{ii}}]/H$\alpha$}}
\newcommand{\NS}{[\mbox{N{\sc{ii}}]/}[S{\sc{ii}}]}
\newcommand{\SII}{[\mbox{S{\sc{ii}}]}}
\newcommand{\OII}{[\mbox{O{\sc{ii}}]}}
\newcommand{\NO}{\mbox{$\log_{10}$(N/O)}}
\newcommand{\OH}{\mbox{$\log_{10}$(O/H)}}
\newcommand{\MT}{M$_{13}$}
\newcommand{\DS}{D$_{16}${}}
\newcommand{\AP}{P$_{\rm 22}${}}

\newcommand{\SIIR}{\SII$\lambda$6716/$\lambda$6731}
\newcommand{\OIIR}{\OII$\lambda$3726/$\lambda$3729}
\newcommand{\nel}{$n_{\rm e}$}

\newcommand{\KTD}{{KMOS$^{\rm 3D}$}}
\newcommand{\magphys}{{\sc{magphys}}}
\newcommand{\galfit}{{\sc{galfit}}}

\newcommand{\sex}{{\sc{sextractor}}}
\newcommand{\eagle}{{\sc{eagle}}}

\newcommand{\zest}[1]{$z\approx#1$}
\newcommand{\zra}[2]{$z\approx#1-#2$}
\newcommand{\Rh}{$R_{\rm h}$}
\newcommand{\Mstar}{$M_{\rm *}$}
\newcommand{\med}[3]{\mbox{#1 = #2\,$\pm$\,#3}}
\newcommand{\zgrad}{$\Delta$Z/$\Delta$R}

\newcommand{\col}[1]{{\color{black}#1\color{black}{}}}
\newcommand{\coln}[1]{{\color{black}#1\color{black}{}}}
\newcommand{\sig}{$\sigma_{\rm H\alpha, 2.2R_d}$}
\newcommand{\vsig}{$v_{\rm H\alpha}/\sigma_{\rm H\alpha}$}

\title[Resolved ISM Properties within $z \approx 1.5$ SFGs]{The resolved chemical abundance properties within the interstellar medium of star-forming galaxies at $\mathbf{ \textit{z} \approx 1.5}$}

\author[S. Gillman et al.]{S. Gillman\orcid{0000-0001-9885-4589},$^{\hyperlink{DAWN}{1},\hyperlink{DTU}{2}}$\thanks{E-mail: srigi@space.dtu.dk}
A. Puglisi\orcid{0000-0001-9369-1805},$^{\hyperlink{CEA}{3}}$  
U. Dudzevi\v{c}i\={u}t\.{e}\orcid{0000-0003-4748-0681},$^{\hyperlink{CEA}{3},\hyperlink{MPE}{4}}$
A. M. Swinbank\orcid{0000-0003-1192-5837},$^{\hyperlink{CEA}{3}}$
A. L.  Tiley\orcid{0000-0002-0617-9510},$^{\hyperlink{CEA}{3},\hyperlink{ICRAR}{5}}$ 
\and 
C. M. Harrison\orcid{0000-0001-8618-4223},$^{\hyperlink{Newcastle}{6}}$
J. Molina\orcid{0000-0002-8136-8127},$^{\hyperlink{China}{7}}$
R. M. Sharples\orcid{0000-0003-3449-8583},$^{\hyperlink{CFAI}{8}}$
R. G. Bower\orcid{0000-0002-5215-6010},$^{\hyperlink{ICC}{9}}$
M. Cirasuolo\orcid{0000-0003-1644-8881},$^{\hyperlink{ESO}{10}}$
\and
Edo Ibar,$^{\hyperlink{Chile}{11}}$
and D. Obreschkow\orcid{0000-0002-1527-0762},$^{\hyperlink{ICAR}{5},\hyperlink{ARC2}{12}}$
\\
$^{1}$\hypertarget{DAWN}{Cosmic Dawn Center (DAWN)}\\
$^{2}$\hypertarget{DTU}{DTU-Space, Technical University of Denmark, Elektrovej 327, DK-2800 Kgs. Lyngby, Denmark}\\
$^{3}$\hypertarget{CEA}{Centre for Extragalactic Astronomy, Department of Physics, Durham University, South Road, Durham, DH1 3LE UK}\\
$^{4}$\hypertarget{MPE}{Max-Planck-Institut für Astronomie, Königstuhl 17, D-69117, Heidelberg, Germany}\\
$^{5}$\hypertarget{ICRAR}{International Centre for Radio Astronomy Research, University of Western Australia, 35 Stirling Highway, Crawley, WA, Australia}\\
$^{6}$\hypertarget{Newcastle}{School of Mathematics, Statistics and Physics, Newcastle University, Newcastle upon Tyne NE1 7RU, UK}\\
$^{7}$\hypertarget{China}{Kavli Institute for Astronomy and Astrophysics, Peking University, Beijing 100871, China} \\ 
$^{8}$\hypertarget{CFAI}{Centre for Advanced Instrumentation, Department of Physics, Durham University, South Road, Durham DH1 3LE UK}\\
$^{9}$\hypertarget{ICC}{Institute for Computational Cosmology, Department of Physics, Durham University, South Road, Durham DH1 3LE UK}\\
$^{10}$\hypertarget{ESO}{European Southern Observatory, Karl-Schwarzschild-Str 2, D-86748 Garching b. München, Germany} \\
$^{11}$\hypertarget{Chile}{Instituto de F\'isica y Astronom\'ia, Universidad de Valpara\'iso, Avda. Gran Breta\~na 1111, Valpara\'iso, Chile}\\
$^{12}$\hypertarget{ARC2}{Australian Research Council Centre of Excellence for All-Sky Astrophysics, 44 Rosehill Street Redfern, NSW 2016, Australia} \\
}

\date{Accepted XXX. Received YYY; in original form ZZZ}

\pubyear{2021}

\begin{document}
\label{firstpage}
\pagerange{\pageref{firstpage}--\pageref{lastpage}}
\maketitle

\begin{abstract}
We exploit the unprecedented depth of integral field data from the KMOS Ultra-deep Rotational Velocity Survey (KURVS) to analyse the strong (H$\alpha$) and forbidden ([N{\sc{ii}}], [S{\sc{ii}]}) emission line ratios in 22 main-sequence galaxies at $z\,\approx\,1.5$.
Using the [N{\sc{ii}}]/H$\alpha$ emission-line ratio we confirm the presence of the stellar mass\,--\,gas-phase metallicity relation at this epoch, with galaxies exhibiting on average 0.13\,$\pm$\,0.04\,dex lower gas-phase metallicity (12+log(O/H)$_{\rm M13}$\,=\,8.40\,$\pm$\,0.03) for a given stellar mass ($\log_{10}$($M_{\rm *}$[$M_{\odot}$]\,=\,10.1\,$\pm$\,0.1) than local main-sequence galaxies. We determine the galaxy-integrated [S{\sc{ii}}] doublet ratio,  with a median value of [S{\sc{ii}}]$\lambda$6716/$\lambda$6731\,=\,1.26\,$\pm$\,0.14 equivalent to an electron density of log$_{10}$($n_{\rm e}$[cm$^{-3}$])\,=\,1.95\,$\pm$\,0.12. Utilising CANDELS \textit{HST} multi-band imaging we define the pixel surface-mass and star-formation rate density in each galaxy and spatially resolve the fundamental metallicity relation at $z\,\approx\,1.5$, finding an evolution of 0.05\,$\pm$\,0.01\,dex compared to the local relation. We quantify the intrinsic gas-phase metallicity gradient within the galaxies using the [N{\sc{ii}}]/H$\alpha$ calibration, finding a median annuli-based gradient of $\Delta$Z/$\Delta$R\,=\,$-$0.015\,$\pm$\,0.005 dex\,kpc$^{-1}$.  Finally we examine the azimuthal variations in gas-phase metallicity, which show a negative correlation with the galaxy integrated star-formation rate surface density ($r_{\rm s}\,$\,=\,$-$0.40,\,$p_{\rm s}$\,=\,0.07) but no connection to the galaxies kinematic or morphological properties nor radial variations in stellar mass surface density or star formation rate surface density. This suggests 
both the radial and azimuthal variations in interstellar medium properties are connected to the galaxy integrated density of recent star formation. 
\end{abstract}

\begin{keywords}
galaxies:abundances -- galaxies:high-redshift -- galaxies:ISM
\end{keywords}

%%%%%%%%%%%%%%%%%%%%%%%%%%%%%%%%%%%%%%%%%%%%%%%%%%

%%%%%%%%%%%%%%%%% BODY OF PAPER %%%%%%%%%%%%%%%%%%

\section{Introduction}

The rest-frame optical morphology of main-sequence star-forming galaxies has evolved over cosmic time. From a clumpy, irregular high-redshift (\zest{1-2}) population to the grand-design Hubble-type 
galaxies seen in the local Universe \cite[e.g.][]{Glazebrook1995,Abraham1996,Ilbert2010,Conselice2014,Sachdeva2019}. This evolution has been defined in terms of the galaxy's fundamental properties (e.g. star-formation rate,  gas fraction, angular momentum and chemical abundance) with high-redshift galaxies exhibiting higher star-formation rates and gas fractions whilst having lower angular momentum and gas-phase metallicity at fixed stellar mass \citep[e.g.][]{Schreiber2009, Whitaker2012, Tacconi2013,Swinbank2017,Gillman2021}. Constraining the physical properties of the interstellar medium, and the role of the evolving baryon cycle, in high-redshift main-sequence galaxies is vital in order to fully explain the nature of these differences \citep[see][for full reviews]{Maiolino2019,Tacconi2020,Schreiber2020}.

Observational studies have shown that the empirical scaling relations present in the local Universe that interconnect a galaxy's fundamental properties are in place at high-redshift, but with variable normalisation and or intrinsic scatter (e.g. the star-formation main-sequence; \citealt{Whitaker2012,Schreiber2015}, angular momentum stellar mass relation; \citealt{Burkert2016,Harrison2017}, Tully-Fisher relation; \citealt{Ubler2017,Tiley2019}). In particular the gas-phase metallicity stellar mass relation (MZR) has been observed up to a redshift of \zest{3.3}, with a normalisation that decreases with increasing redshift, at fixed stellar mass \citep[e.g.][]{Mannucci2010,Yabe2015,Steidel2014,Wuyts2016,Kashino2017,Sanders2020,Gillman2021}.

The gas-phase metallicity of the interstellar medium in distant galaxies is quantified using strong-line calibrations of rest-frame optical emission lines \citep[e.g.][]{Pettini2004,Kewley2008,Marino2013,Dopita2016,Poetrodjojo2021}. By using these locally derived  calibrations, that convert an emission-line ratio (e.g. \NH{}, O$_{\rm 3}$N$_{\rm 2}$, R$_{\rm 23}$) to an Oxygen abundance (12+\OH{}), the gas-phase metallicity in distant galaxies can be constrained. The decrease in normalisation of the MZR at higher redshifts is attributed to the higher star-formation rates and increased metal-poor gas inflows at earlier cosmic times as well as evolution in the outflow strength, gas fraction and star-formation efficiency of galaxies \citep[e.g.][]{Lilly2013, Lian2018,Sanders2020}. Tracing the distribution of metallicity in galaxies across cosmic time provides empirical constraints for theoretical models aimed at describing the complex interplay between star formation, gas flows and feedback processes that drive a galaxy's evolution \citep[e.g.][]{Trayford2019,Belfiore2019,Sharda2021,Yates2021}. 

However, a number of physical parameters appear to determine the strong optical emission-line fluxes produced in \HII{} regions. These include the chemical abundance of the gas but also the shape and normalisation of the ionising spectrum, the ionisation state of the gas and the gas density \citep[e.g.][]{Dopita1986,Kewley2002,Dopita2006a,Dopita2006b,Kewley2019b,Kumari2021,Ji2021J,Curti2021,Helton2022}. These strong-line calibrations assume that the interstellar medium conditions at high-redshift are similar to those found in local galaxies, such that the variation in emission-line ratios is driven purely by changes in the chemical abundance \citep[e.g.][]{Bian2018,Bian2020,Hayden-Pawson2021}. By defining the fundamental properties of the interstellar medium in distant galaxies, we can begin to constrain the origin of the variation in emission-line ratios.

In the local Universe spatially-resolved studies of galaxies have identified negative metallicity gradients in star-forming galaxies \citep[e.g.][]{Sanchez2014,Belfiore2017,Lutz2021} as well as the presence of the resolved mass - metallicity relation on sub-kpc scales \citep[e.g.][]{Barrera-Ballesteros2016,Erroz-Ferrer2019,Sanchez2020,Neumann2021}. The secular evolution of local galaxies is well prescribed by `inside-out' models of galaxy evolution that explain the higher metallicity in the central regions \citep[e.g.][]{Nelson2012}. 

In addition to robust measures of the chemical abundance properties, constraining the emission-line ratios of faint forbidden emission-lines (e.g. \NII{}, \SII{}) in galaxies enables constraints on other interstellar medium properties such as electron density \citep[e.g.][]{Dopita2016,Maiolino2019,Kewley2019b,Kewley2019a}. In distant galaxies however, measuring even integrated properties of faint emission-lines is challenging due to the low-surface brightness and signal to noise (S/N). Previous studies have overcome these issues by stacking galaxy spectra  in order to measure sample averaged emission-line properties \citep[e.g.][]{Swinbank2019,Schreiber2019, Davies2021,Alaina2021} or utilised either space-based low-resolution grism spectroscopy \citep[e.g.][]{Jones2015,Wang2020,Simons2020,Backhaus2021} or gravitationally lensed galaxies \citep[e.g.][]{Curti2019,Patrico2019,Florian2020,Hayden-Pawson2021}.

To define the interstellar medium properties of high-redshift galaxies and establish whether the global scaling relations (e.g. mass\,--\,metallicity, star-formation rate main\,--\,sequence) originate from localised relations within individual galaxy's requires us to go beyond stacks and even integrated emission-line ratios for individual distant galaxies. 
To this end, in this paper we present an analysis of the chemical abundance properties of the interstellar medium, using a variety of metallicity indicators, in 22 galaxies from the KMOS Ultra-deep Rotational Velocity Survey (KURVS; Puglisi et al. in preparation, hereafter \AP) each observed for $\approx$70 hours on source with KMOS. 

In Section \ref{Sec:MSFG} we present the observations and the main-sequence properties of KURVS galaxies. We then define the gas-phase metallicity properties and emission-line ratios in Section \ref{Sec:metals} before quantifying the electron density in Section \ref{Sec:edensity}. In Section \ref{Sec:resolved_FMR} we derive the resolved fundamental metallicity relation before finally in section \ref{Sec:profiles} quantifying the radial profiles of metallicity and electron density before giving our conclusions in Section \ref{Sec:Conc}.

A Nine-Year Wilkinson Microwave Anisotropy Probe \citep{Hinshaw2013} cosmology is used throughout this work with $\Omega_{\Lambda}$\,=\,0.721, $\Omega_{\rm m}$\,=\,0.279 and H$_{\rm 0}$\,=\,70\,km\,s$^{-1}$ Mpc$^{-1}$.
In this cosmology a spatial resolution of 0.57 arcsecond (the median FWHM of the seeing in our data) corresponds to a physical scale of 4.9\,kpc at a redshift of \zest{1.5}. All quoted magnitudes are on the AB system and stellar masses are calculated assuming a Chabrier initial mass function (IMF) \citep{Chabrier2003}.

\begin{figure*}
    \centering
    \includegraphics[width=\linewidth]{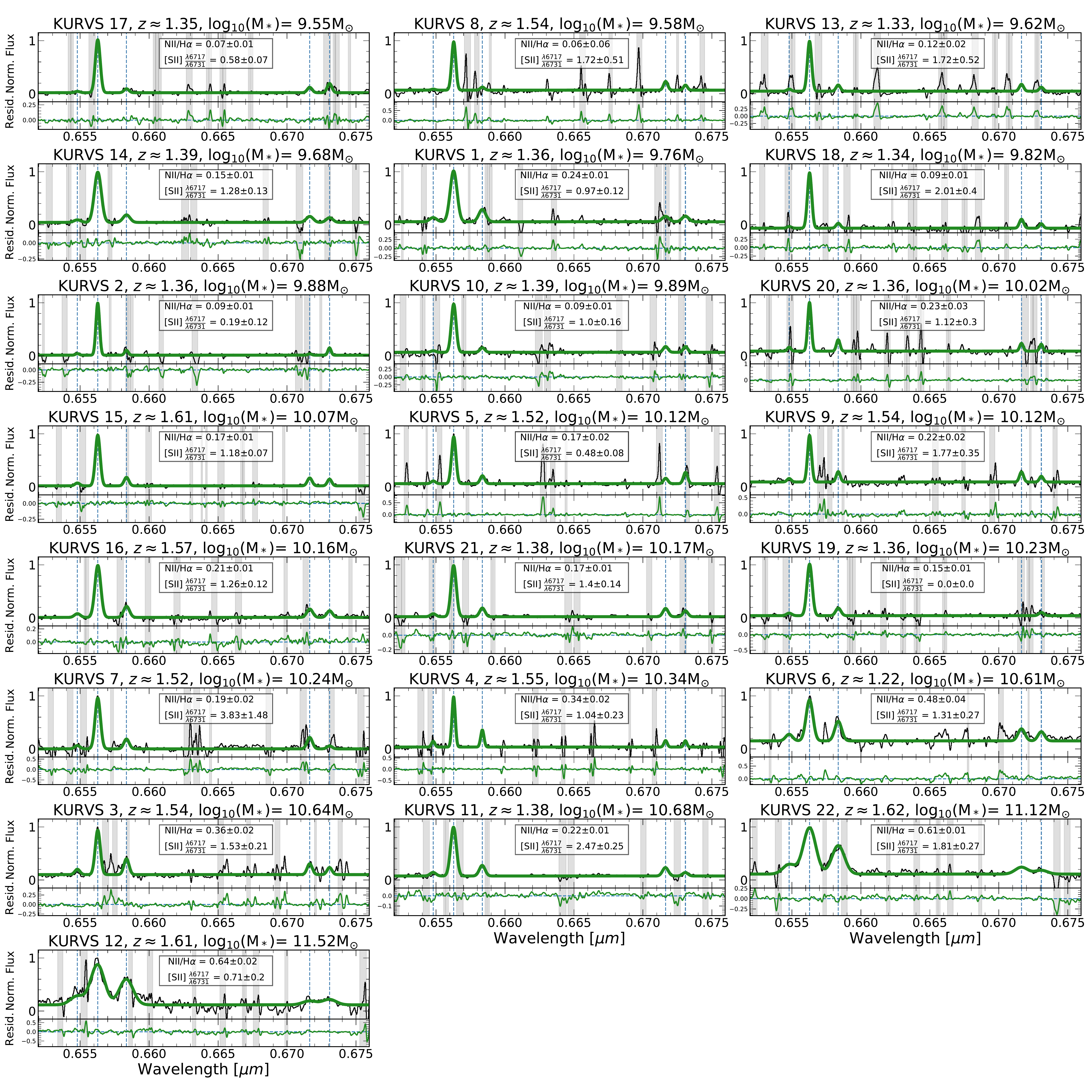}
    \caption{The de-redshifted $H$-band rest-frame integrated KMOS spectra from the KURVS-CDFS sample ranked by stellar mass. We highlight the \HA{}, \NII{} and \SII{} emission lines as well as overlaying the combined Gaussian profile fit  (and residual) from which we measure the \NH{}, \SII-ratio and \SII/\HA{} emission-line  ratios as shown in Table \ref{Table:KURVS_metals} and each panel. We highlight (in grey) the strong OH emission lines present in each spectrum. For each galaxy we also label its ID, spectroscopic \HA{} redshift, \magphys{} derived stellar mass, galaxy-integrated \NH{} and \SIIR{} ratio.}
    \label{fig:SII_spec}
\end{figure*}

\begin{table*} 
\centering 
\caption{Main sequence and emission-line properties of the KURVS-CDFS sample.}
\begin{tabular}{lllllllllll}
\hline
ID & RA($^\circ$) & DEC($^\circ$)  &$z_{\rm spec, H\alpha}$ & $\rm \log_{10}{M_*^{\rm a}}$ [M$_{\odot}$] & SFR$^{\rm b}_{\rm H\alpha}$ [M$_{\odot}$yr$^{-1}$] & R$_{\rm h}$ [kpc]$^{\rm c}$ & \NH$^{\rm d}$ & \SH$^{\rm e}$ & \NII/\SII$^{\rm f}$ & \SII $\frac{\lambda 6717}{\lambda 6731}$ \\
\hline
KURVS 1 & 53.0723 & -27.8980 & 1.3591 & 9.76 & 30 $\pm$ 1 & 1.08 $\pm$ 0.05 & 0.24 $\pm$ 0.01 & 0.11 $\pm$ 0.01 & 2.18 $\pm$ 0.19 & 0.97 $\pm$ 0.12 \\
KURVS 2 & 53.0625 & -27.8840 & 1.3599 & 9.88 & 3 $\pm$ 0 & 8.04 $\pm$ 0.38 & 0.09 $\pm$ 0.01 & $-^{\ddagger}$ & $-$ & $-$ \\
KURVS 3 & 53.0921 & -27.8792 & 1.5411 & 10.64 & 11 $\pm$ 2 & 4.20 $\pm$ 0.08 & 0.36 $\pm$ 0.02 & 0.16 $\pm$ 0.02 & 2.22 $\pm$ 0.30 & 1.53 $\pm$ 0.21 \\
KURVS 4 & 53.0841 & -27.8773 & 1.5525 & 10.34 & 20 $\pm$ 1 & 3.81 $\pm$ 0.07 & 0.34 $\pm$ 0.02 & 0.12 $\pm$ 0.02 & 2.73 $\pm$ 0.55 & 1.04 $\pm$ 0.23 \\
KURVS 5 & 53.0574 & -27.8674 & 1.5183 & 10.12 & 13 $\pm$ 1 & 3.17 $\pm$ 0.04 & 0.17 $\pm$ 0.02 & 0.24 $\pm$ 0.02 & 0.70 $\pm$ 0.09 & 0.48 $\pm$ 0.08 \\
KURVS 6 & 53.0585 & -27.8568 & 1.2213 & 10.61 & 6 $\pm$ 1 & 2.90 $\pm$ 0.04 & 0.48 $\pm$ 0.04 & 0.22 $\pm$ 0.04 & 2.14 $\pm$ 0.47 & 1.31 $\pm$ 0.27 \\
KURVS 7 & 53.0468 & -27.8520 & 1.5183 & 10.24 & 6 $\pm$ 0 & 4.51 $\pm$ 0.08 & 0.19 $\pm$ 0.02 & $-^{\ddagger}$ & $-$ & $-$ \\
KURVS 8 & 53.0285 & -27.8487 & 1.5394 & 9.58 & 7 $\pm$ 1 & 2.50 $\pm$ 0.19 & $\downarrow$0.06$^{\dagger}$ & 0.11 $\pm$ 0.03 & 0.61 $\pm$ 0.27 & 1.72 $\pm$ 0.51 \\
KURVS 9 & 53.0897 & -27.8446 & 1.5402 & 10.12 & 3 $\pm$ 0 & 5.03 $\pm$ 0.08 & 0.22 $\pm$ 0.02 & 0.12 $\pm$ 0.02 & 1.81 $\pm$ 0.40 & 1.77 $\pm$ 0.35 \\
KURVS 10 & 53.1247 & -27.8422 & 1.3899 & 9.89 & 3 $\pm$ 0 & 1.90 $\pm$ 0.05 & 0.09 $\pm$ 0.01 & 0.12 $\pm$ 0.01 & 0.79 $\pm$ 0.15 & 1.00 $\pm$ 0.16 \\
KURVS 11 & 53.1418 & -27.8413 & 1.3837 & 10.68 & 65 $\pm$ 3 & 5.94 $\pm$ 0.03 & 0.22 $\pm$ 0.01 & 0.07 $\pm$ 0.01 & 3.11 $\pm$ 0.31 & 2.47 $\pm$ 0.25 \\
KURVS 12 & 53.1314 & -27.8413 & 1.6130 & 11.52 & 19 $\pm$ 2 & 6.52 $\pm$ 0.07 & 0.64 $\pm$ 0.02 & 0.14 $\pm$ 0.02 & 4.69 $\pm$ 0.83 & 0.71 $\pm$ 0.20 \\
KURVS 13 & 53.0521 & -27.8391 & 1.3338 & 9.62 & 14 $\pm$ 1 & 3.41 $\pm$ 0.04 & 0.12 $\pm$ 0.02 & 0.08 $\pm$ 0.02 & 1.54 $\pm$ 1.09 & 1.72 $\pm$ 0.52 \\
KURVS 14 & 53.1557 & -27.8371 & 1.3893 & 9.68 & 11 $\pm$ 0 & 2.55 $\pm$ 0.04 & 0.15 $\pm$ 0.01 & 0.09 $\pm$ 0.01 & 1.60 $\pm$ 0.15 & 1.28 $\pm$ 0.13 \\
KURVS 15 & 53.0706 & -27.8345 & 1.6134 & 10.07 & 28 $\pm$ 1 & 3.75 $\pm$ 0.05 & 0.17 $\pm$ 0.01 & 0.13 $\pm$ 0.01 & 1.27 $\pm$ 0.07 & 1.18 $\pm$ 0.07 \\
KURVS 16 & 53.1546 & -27.8280 & 1.5691 & 10.16 & 18 $\pm$ 1 & 3.93 $\pm$ 0.04 & 0.21 $\pm$ 0.01 & 0.13 $\pm$ 0.01 & 1.66 $\pm$ 0.15 & 1.26 $\pm$ 0.12 \\
KURVS 17 & 53.0463 & -27.8273 & 1.3528 & 9.55 & 7 $\pm$ 0 & 2.43 $\pm$ 0.04 & 0.07 $\pm$ 0.01 & 0.17 $\pm$ 0.01 & 0.43 $\pm$ 0.06 & 0.58 $\pm$ 0.07 \\
KURVS 18 & 53.1360 & -27.8135 & 1.3416 & 9.82 & 6 $\pm$ 0 & 3.21 $\pm$ 0.07 & 0.09 $\pm$ 0.01 & 0.08 $\pm$ 0.01 & 1.11 $\pm$ 0.30 & 2.01 $\pm$ 0.40 \\
KURVS 19 & 53.0825 & -27.8109 & 1.3573 & 10.23 & 9 $\pm$ 0 & 2.07 $\pm$ 0.03 & 0.15 $\pm$ 0.01 & $-^{\ddagger}$ & $-$ & $-$ \\
KURVS 20 & 53.1043 & -27.7884 & 1.3563 & 10.02 & 21 $\pm$ 1 & 1.20 $\pm$ 0.02 & 0.23 $\pm$ 0.03 & 0.14 $\pm$ 0.03 & 1.64 $\pm$ 0.48 & 1.12 $\pm$ 0.30 \\
KURVS 21 & 53.1310 & -27.8604 & 1.3822 & 10.17 & 15 $\pm$ 1 & 2.85 $\pm$ 0.05 & 0.17 $\pm$ 0.01 & 0.12 $\pm$ 0.01 & 1.46 $\pm$ 0.15 & 1.40 $\pm$ 0.14 \\
KURVS 22 & 53.1572 & -27.8334 & 1.6180 & 11.12 & 32 $\pm$ 1 & 4.10 $\pm$ 0.04 & 0.61 $\pm$ 0.01 & 0.08 $\pm$ 0.01 & 7.46 $\pm$ 1.03 & 1.81 $\pm$ 0.27 \\
\hline\\[-0.5cm]  
\end{tabular} 
\label{Table:KURVS_metals} 
\begin{flushleft}
\footnotesize{
\hspace{0.cm}  a) Homogeneous uncertainty of $\pm$\,0.2 dex to account for the uncertainties in \magphys{} derived stellar mass \citep{Mobasher2015}.\\
\hspace{0.cm}  b) \HA{} SFR corrected for extinction
\hspace{0.5cm} c) 1.6$\mu$m half-light radius
\hspace{0.5cm} d) \NII$\lambda$6584 / H$\alpha$ 
\hspace{0.5cm} e) \SII$\lambda$6717+$\lambda$6731 / H$\alpha$ \\
\hspace{0.cm} f) \NII$\lambda$6584 / \SII$\lambda$6717+$\lambda$6731
\hspace{0.5cm} $^+=$ AGN candidate (see Section \ref{Sec:AGN})
\hspace{0.5cm} $\dagger$\,=\,$2\sigma$ upper limit on \NII{} 
\hspace{0.5cm} $\ddagger$\,=\, S/N \SII{}$<$2
}
\end{flushleft}
\end{table*}

\section{Observations, Sample $\&$ Galaxy Integrated Properties}\label{Sec:MSFG}

The 44 galaxies in the KURVS sample (see \AP) are $H$-band selected main-sequence galaxies in the redshift range \zra{1.25}{1.75} with existing H$\alpha$ detections (F$\rm_{H\alpha,obs}\geq 5 \times 10^{-17}\,ergs^{-1}\,cm^{-2}$). In this paper we concentrate on the first half of the sample (KURVS-CDFS), which comprises sources in the CDFS field for which observations are complete. 

The majority of galaxies (20/22) have previously been observed as part of the KMOS Galaxy Evolution Survey \citep[KGES;][]{Gillman2019,Tiley2021} with two galaxies (KURVS 21 and 22) selected from the \KTD{} survey \citep{Wisnioski2019}. The KURVS galaxies lie within the Cosmic Assembly Near-infrared Deep Extragalactic Legacy Survey \citep[CANDELS;][]{Grogin2011} \textit{HST} fields. For each galaxy there is a wealth of ancillary multi-wavelength imaging data from the ultraviolet to mid-infrared from which we can construct the spectral energy distribution (SED) of the each galaxy. In this section we define the KMOS observations and quantify the main-sequence properties of the KURVS galaxies.

\subsection{KMOS Observations}\label{Sec:Spec}

All the galaxies in the KURVS sample were observed using KMOS, a multi-object spectrograph mounted on the Nasmyth focus of the 8-m class UT1 telescope at the VLT, Chile. The spectrograph has 24 individual integral-field units that patrol a 7.2 arcmin diameter field, each with a 2.8 × 2.8 arcsec$^2$ field of view and 0.2 × 0.2 arcsec$^2$ spaxels.

The KMOS observations cover the rest-frame near-infrared $H$-band spectrum from 1.45\,--\,1.87$\mu$m and equating to $\approx$70h on-source per galaxy taken between October 2018 - December 2019. 
The average full width half maximum (FWHM) of the seeing of the observations is $\approx$\,0.57 arcseconds, as determined from the dedicated PSF star frames in the KMOS mask. Full details of the observations and data reduction are discussed in \AP. In brief, the data were reduced using the European Southern Observatory (ESO) Recipe Execution Tool \citep{ESO2015} which extracts, wavelength calibrates and flat fields each of the spectra and forms a data cube from each observation. 

In addition to the KURVS sample, throughout this paper we include observations from the KGES parent sample from which the KURVS galaxies were selected. The KGES survey is a KMOS large program of 288 $H-$band selected star-forming galaxies at \zra{1.25}{1.75}. Of those with \HA{} detections (243/288) the median stellar mass and dust-corrected \HA{} star-formation rate is \med{$\log_{10}$(\Mstar[$M_{\odot}$])}{10.1}{0.1}, \med{SFR$\rm_{H\alpha}$[M$_{\odot}$yr$^{-1}$]}{17}{2} respectively. The emission-line ratios of the KGES galaxies \citep[see][]{Gillman2021} used throughout this paper are derived using consistent methodology to our analysis of the KURVS sample.

For our parent sample comparison analysis, we select galaxies from the KGES survey with a galaxy-integrated \NII{} and \SII{} S/N$>$3 to ensure robust measurements of their integrated gas-phase metallicity and electron density. In total the KGES sample used in this paper comprises of 71 galaxies with a  medium redshift (and bootstrapped uncertainty) of  \med{$z$}{1.48}{0.01}. The KGES sub-sample comprises of main-sequence galaxies with a median stellar mass of \med{$\log_{10}$(\Mstar[$M_{\odot}$])}{10.1}{0.1} and \HA-derived extinction-corrected star-formation rate of \med{SFR$\rm _{H\alpha}$[M$_{\odot}$yr$^{-1}$]}{23}{3}, comparable to the full KGES \HA{} detected sample, with a preference towards higher star-formation rate galaxies, by selection.

\subsection{Galaxy Spectra $\&$ AGN Identification} \label{Sec:AGN}
At \zest{1.5} the rest-frame optical nebula emission lines (e.g \HA, \NII, and \SII{}) are redshifted into the near-infrared $H$-band probed by KMOS. These emission lines, alongside a number of atmospheric OH airglow emission lines, are visible in the KMOS spectra. In Figure \ref{fig:SII_spec} we show the integrated spectra for the 22 galaxies in the KURVS sample, ranked by their stellar mass. We highlight (in grey) the strong OH emission lines present in each spectrum, and overlay a Gaussian profile fit (in green) to the \HA, \NII, and \SII{} emission lines as well as showing the residuals. Each data cube has been de-redshifted and the velocity field removed. The integrated spectrum is a sum of the spaxels resolved in the \HA{} velocity map of the galaxy, as discussed in Section \ref{Sec:metals}.

Figure \ref{fig:SII_spec} demonstrates the presence of the stellar mass -- metallicity relation in the KURVS sample, with the increasing presence of forbidden emission lines with increasing stellar mass. It is important to understand whether the emission lines shown in Figure \ref{fig:SII_spec} originate from star formation or active galactic nuclei (AGN).
The presence of an AGN would enhance the emission-line ratios of a galaxy and thus make them a unreliable tracer of the galaxies metallicity.

To determine the presence of AGN in the KURVS sample, we first utilise the AGN identification scheme employed by \citet{Tiley2021} for the full KGES sample. In short AGN are identified in our sample by criteria placed on their integrated \NH{} ratio, emission-line width and infrared colours following the  \citet{Donley2012} and \citet{Stern2012} colour selection (see \citet{Tiley2021} for full details). We also exclude galaxies with bright X-ray counterparts by cross-matching our sample with the  CDFS 7 Ms Source Catalog \citep{Luo2017}.

Of the 22 galaxies in the KURVS sample, we identify two candidates where an AGN may influence the line ratios. KURVS 11 and KURVS 12 have $WISE$\footnote{The All WISE Source Catalog \citep{Cutri2013}} colour W1[3.6]$-$W2[4.5]$>$0.8, whilst KURVS 12 is also highlighted in the \citet{Luo2017} catalogue as an AGN candidate with an intrinsic 0.5\,--\,7.0\,kev luminosity of  $L_{\rm X,int}$=5.045$\times$10$^{42}$ ergs$^{-1}$. KURVS 11 has an intrinsic 0.5\,--\,7.0\,kev X-ray luminosity of $L_{\rm X,int}$=5.312$\times$10$^{41}$ ergs$^{-1}$, but this is below the AGN threshold of the \citet{Luo2017} catalogue ($L_{\rm X,int}\geq$3$\times$10$^{42}$ ergs$^{-1}$). Three other galaxies in the sample (KURVS 14, 15 and 22) have X-ray counterparts but luminosities that lie below the AGN threshold. None of the galaxies in the sample have a \NH{}$\geq$0.8 or a $\rm \sigma_{H\alpha}\geq 1000\,kms^{-1}$ in their integrated spectrum from a 1.2 arcsecond aperture.

In the following analysis we highlight KURVS 11 and 12 as candidate AGN in the relevant plots but do not omit them from our analysis. In total we have 20 star-forming main-sequence galaxies and 2 potential AGN. Having established the origin of the emission in Figure \ref{fig:SII_spec}, we now focus on the main-sequence properties of the KURVS galaxies to establish whether they are typical star-forming galaxies at \zest{1.5}.

\begin{figure*}
\centering
\begin{subfigure}{.5\textwidth}
  \centering
  \includegraphics[width=\linewidth]{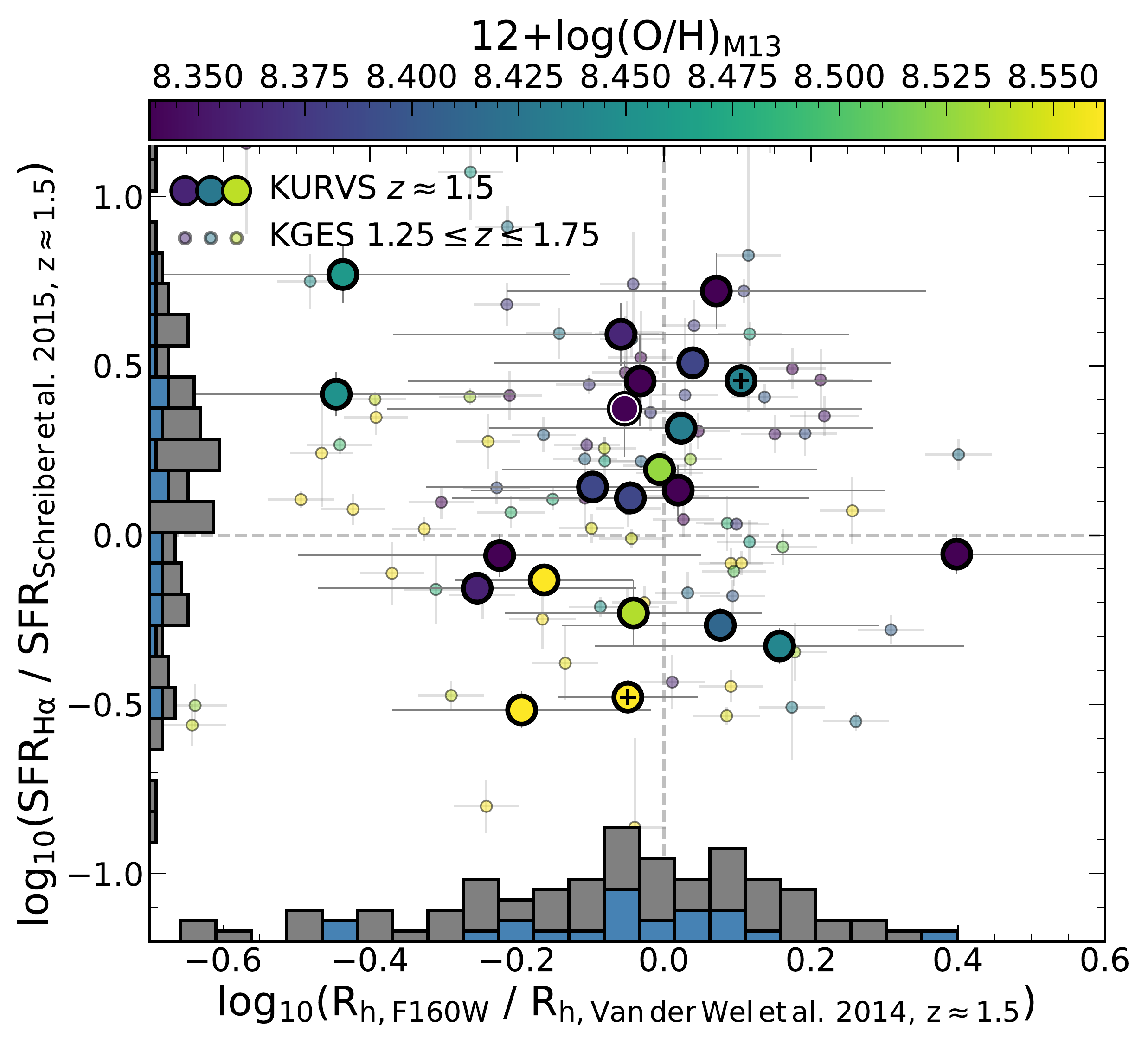}
\end{subfigure}%
\begin{subfigure}{.5\textwidth}
  \centering
  \includegraphics[width=\linewidth]{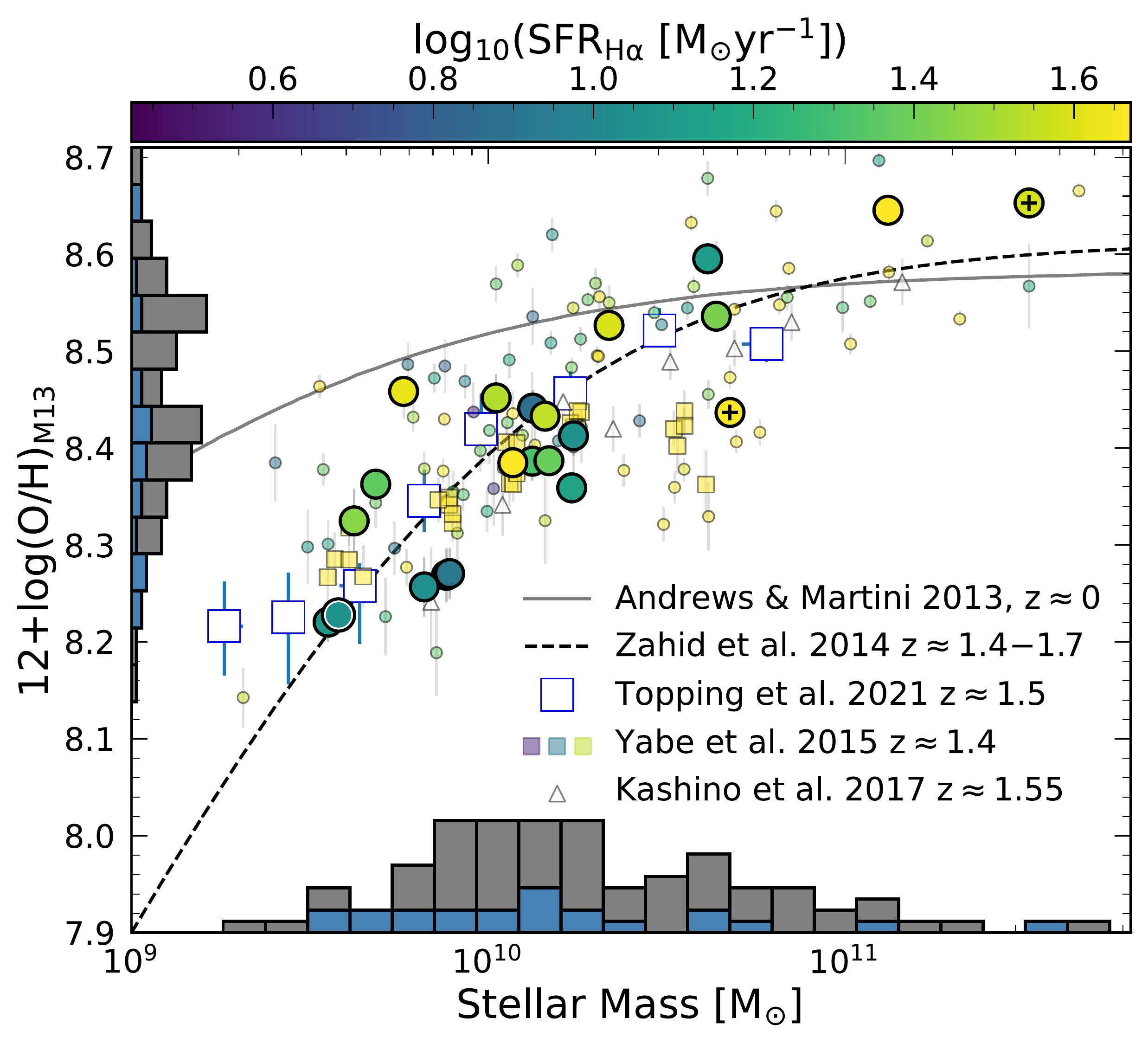}
\end{subfigure}
\caption{\textbf{Left:} The offset from the star-formation rate $-$ stellar mass plane quantified by \citet{Schreiber2015} as a function of the offset from the stellar continuum half-light radii $-$ stellar mass plane defined by \citet{VanderWel2014} at \zest{1.5}. The KURVS-CDFS star-forming galaxies (\textit{circles}), AGN (\textit{circles + cross}; see Section \ref{Sec:AGN}) and KGES galaxies with F160W \textit{HST} data (\textit{dots}) are coloured by their integrated gas-phase metallicity derived using the \NH{} ratio following the \citet{Marino2013} calibration (Section \ref{Sec:metals}). A histogram for each sample is shown on each axis. The KURVS-CDFS galaxies represent main-sequence galaxies at \zest{1.5} with a small scatter about the SFR$-$\Mstar{} and \Rh$-$\Mstar{} relations at this epoch. 
\textbf{Right:} Gas-phase mass\,--\,metallicity relation for the KURVS-CDFS (\textit{circles}) and KGES (\textit{dots}) galaxies. We also show two comparison samples from the \zest{1.6} \citet{Kashino2017} FMOS-COSMOS sample ($triangles$) and the \citet{Yabe2015}  Subarau-FMOS sample ($squares$) at \zest{1.4}.
The grey black line represents the mass\,--\,metallicity relation at $z$\,$\approx$\,0 derived from Sloan Digital Sky Survey (SDSS) \citep{Andrews2013} whilst the dashed black line indicates the mass\,--\,metallicity relation at \zra{1.4}{1.7} from the Fiber-Multi Object Spectrograph (FMOS) survey \citep{Zahid2014}.
We also show the results from \citet{Topping2021} for the stacked spectra of 30 galaxies from the MOSDEF Survey at \zest{1.5} (blue squares). 
All comparison samples are converted to an Oxygen abundance derived from the \citet{Marino2013} \NH{} ratio calibration.
In both panels the white outlined circle indicates a $2-\sigma$ limit derived on the \NII{} line for objects with low S/N (see Table \ref{Table:KURVS_metals} and Section \ref{Sec:metals}). KURVS-CDFS galaxies occupy a similar parameter space to other studies at this epoch, with a lower gas-phase metallicity for a given stellar mass compared to \zest{0}.}
\label{fig:MS_MZR}
\end{figure*}

\subsection{Main Sequence Properties}

To quantity the position of the KURVS-CDFS galaxies relative to the main-sequence population at \zest{1.5} and determine whether they represent typical main-sequence galaxies at this epoch, we first need to measure the stellar mass, star-formation rate and stellar continuum half-light radii of the galaxies. 

\citet{Gillman2020} measure the stellar masses of the KGES galaxies by fitting their UV\,--\,mid-infrared SEDs with the \magphys{} software \citep{Cunha2008,daCunha2015}. For the two galaxies selected form the \KTD{} survey, we employ this process, deriving a  median stellar mass (and bootstrapped uncertainty) for the KURVS-CDFS sample of \med{$\log_{10}$(\Mstar[$M_{\odot}$])}{10.1}{0.1}. 

The deep CANDELS \citep{Koekemoer2011} \textit{HST} imaging data of the KURVS-CDFS galaxies can be used to quantify the stellar continuum half-light radius of the galaxies. Using the 1.6$\mu$m (F160W) imaging \citet{VanderWel2014} derived the stellar-continuum size (\Rh{}), S\'ersic index ($n$) and axis ratio of the CANDELS galaxies modelling the two-dimensional stellar-light profiles of the galaxies with the \galfit{} \citep{Galfit2011} software. This analysis was independently verified in \citet{Gillman2020} for the KGES galaxies that lie within the CANDELS \textit{HST} fields. The median F160W stellar continuum half-light radius of the KURVS-CDFS galaxies is \med{\Rh[kpc]}{3.28}{0.42}. This places the KURVS-CDFS sample within 1$-\sigma$ of the expected size (\Rh[kpc]\,=\,3.48) from the \citet{VanderWel2014} mass-size relation at $\log_{10}$(\Mstar[$M_{\odot}$])\,=\,10.1.

The derivation of the \HA{} fluxes for the KURVS-CDFS objects is discussed in \AP. 
The \HA{} flux is corrected for dust extinction using the V-band stellar attenuation ($A_V$) derived from \magphys{} SED fitting and applying the methods of \citet{Wuyts2013}  and  assuming a \citet{Calzetti1994} extinction law \col{such that,
\begin{equation}
    A_{\rm H\alpha,gas} =  A_{\rm H\alpha,stars}(1.9-0.15 A_{\rm H\alpha,stars}),
\end{equation}
as defined in \citet{Stott2016} where  $A_{\rm H\alpha,stars}$ is the rest-frame stellar attenuation at the wavelength of \HA{}.}
The median rest-frame nebular attenuation at the wavelength of \HA{} ($A_{\rm H\alpha,gas}$) for the sample is  $A_{\rm H\alpha,gas}=1.21$. We derive a median \HA{} star-formation of \med{SFR$_{\rm H\alpha}$[M$_{\odot}$yr$^{-1}$]}{24}{6}, which is comparable to the KGES comparison sample. In Table \ref{Table:KURVS_metals} we summarise the main-sequence properties of the KURVS-CDFS sample. 

To establish the position of the KURVS-CDFS galaxies with respect to star-forming galaxy population at \zest{1.5}, we define the relation between stellar mass and star-formation rate (SFR$-$\Mstar) following \citet{Schreiber2015} and the relation between stellar continuum half-light radii and stellar mass (\Rh$-$\Mstar{}) following \citet{VanderWel2014} at \zest{1.5}. In Figure \ref{fig:MS_MZR} we show offset from the SFR$-$\Mstar{} relation  as function of the offset from the median \Rh$-$\Mstar{} for the KURVS-CDFS sample. Galaxies from the KGES comparison sample are shown in the background, with histograms of each sample shown on each axis, indicating a similar distributions of stellar mass and star-formation rate in both the 71 selected KGES galaxies and the KURVS-CDFS sample. The two AGN (as defined in Section \ref{Sec:AGN}) are indicated by crosses and we colour the galaxies by their gas-phase metallicity as derived following the strong-line calibration of \citet{Marino2013} (Section \ref{Sec:metals}). 

Galaxies that exhibit smaller stellar continuum sizes tend to have  lower \HA{} star-formation rates than `typical' galaxies at \zest{1.5} of the same stellar mass. Figure \ref{fig:MS_MZR} indicates that KURVS-CDFS galaxies represent main-sequence galaxies at \zest{1.5} with a small scatter about the SFR$-$\Mstar{} and \Rh$-$\Mstar{} relations at this epoch. By quantifying the properties of the emission lines in the KURVS-CDFS galaxies, as we go on to do in the next section, we can derive physical constraints on the chemical abundance  and electron density properties of the interstellar medium in main-sequence galaxies $\approx$10\,Gyr ago.

\section{Analysis}

In this section we measure the emission-line properties of the KURVS-CDFS galaxies. We first derive the galaxies integrated gas-phase Oxygen abundance, comparing different strong-line calibrations using the \HA, \NII, and \SII{} emission lines and their correlations with galaxy properties. Using the \SIIR{} ratio we then measure the electron density in each galaxy before focusing on the spatially resolved emission-line properties of the sample.

\subsection{Galaxy Integrated Metallicity}\label{Sec:metals}

The metallicity of ionised gas in the interstellar medium, as traced by a particular element can be quantified from the elements emission line flux. For local galaxies this is achieved by directly measuring the electron temperature and density in high signal to noise spectra \citep[see][for full reviews]{Maiolino2019,Kewley2019b}. At high redshift this is not feasible and instead photoionization models are used to predict the relative strength of nebular emission lines. 

To quantify the gas-phase metallicity of the galaxies in the KURVS-CDFS sample, we first de-redshift and remove the velocity field of the galaxies from the datacubes. To do so we normalize each spaxel's spectra by the spectroscopic redshift of the galaxy, as determined from the \HA{} emission line. We then use the \HA{} velocity map of the galaxy, as derived in \AP, to shift each spaxels spectrum by the mean velocity at that position. The result, as shown in Figure \ref{fig:SII_spec}, is that the \HA{} emission line is centred at the rest-frame wavelength of 6563\r{A}.

We note that the velocity field is derived using an adaptive binning technique on the datacube with a spaxel \HA{} signal-to-noise (S/N) threshold of S/N$\geq$5 (see \AP{} for details). This velocity field thus defines the region in which we calculate the de-redshifted integrated spectrum of the galaxy. Any spaxels outside of the velocity map (i.e. S/N$<$5) are excluded as we do not have a velocity correction for these spaxels and we do not expect faint lines beyond the extent of the \HA{} and \NII{} map.
By transforming the spectrum to rest-frame, the rotational broadening of the emission lines are removed. We can then stack the spectra in individual spaxels to measure both the integrated and radial dependence of metallicity in each galaxy.

We apply this procedure to all 22 galaxies in the KURVS-CDFS sample. To quantify the emission-line properties of the KURVS-CDFS galaxies we define an integrated spectra for each galaxy by summing the spectra in the de-redshifted KMOS data cubes. We then fit an emission line model composed of five Gaussian profiles to the \HA{}, \NII{} and \SII{} emission lines present in each galaxies spectra. The Gaussian profile centers  (redshift) and FWHM (velocity dispersion) are fixed to a common value and we set the flux ratio of the \NII{} doublet to be 2.8 following \citet{Osterbrock2006}. To avoid erroneous fits,  the fitting was performed using a $\chi^2$ minimisation method which weights against the wavelengths of the brightest OH skylines (as visible in Figure \ref{fig:SII_spec}).

To ensure robust measurements of interstellar medium properties, we require the S/N of the 6583\r{A} \NII{} emission-line and the weakest \SII{} emission-line to $\geq$2. For galaxies with a \NII{} S/N$<$2, we define a 2\,--\,$\sigma$ limit, whilst for objects with a S/N$<$2 in either \SII{} line  we refrain from deriving interstellar medium properties from the \SII{} lines. In total 21/22 (95$\%$) have an integrated \NII{} S/N$\geq$2\footnote{KURVS 8 has \NII{} S/N$\leq$2} whilst 19/22 (86$\%$) galaxies have an \SII{} S/N$\geq2$\footnote{KURVS 2, 7 and 19 have \SII{} S/N$\leq$2}. For each galaxy we then extract the emission-line ratios as reported in Table \ref{Table:KURVS_metals}. Using these values we can infer the chemical abundance and electron density within each galaxy.

\subsubsection{N2 Index}

A commonly used tracer of gas-phase metallicity at high-redshift is the \NH{} ratio. We use the \NH{} strong-line calibration from \citet{Marino2013} (hereafter \MT) to derive the galaxies gas-phase metallicity. The calibration derived by \MT{} is more accurate at high metallicities (12+log(O/H)$>$8.2) due to the inclusion of spectroscopic observations allowing the electron temperature to be constrained in this regime. This is opposed to the photoionization modelling originally used by \citet{Pettini2004}.

The calibration is defined as;
\begin{equation}\label{Eqn:N2}
    12 + \log(\rm O/H)_{\rm M13} = 8.743 +0.462 × \log_{10}(\NH)
\end{equation}
and has an inherent uncertainty of 0.16\,dex. The calibration is valid for an \NH{} ratio in the range $-1.6 \leq \log_{10}(\NH) \leq -0.2$ or equivalently 12+$\log_{10}$(O/H)$\geq$8. \citet{Poetrodjojo2021} demonstrate that the \MT{} calibration is in closer agreement with other metallicity indicators than the \citet{Pettini2004} calibration when inferring a galaxy's gas-phase Oxygen abundance.

The median \NH{} ratio (and bootstrapped uncertainty) of the KURVS-CDFS sample is \med{\NH}{0.18}{0.03} which corresponds to median metallicity of \med{12+log(O/H)$_{\rm M13}$}{8.40}{0.03}. In Figure \ref{fig:MS_MZR} we show the mass--metallicity relation (MZR) for the KURVS-CDFS sample derived using the \MT{} calibration. We indicate the relations from \citet{Andrews2013} at \zest{0}, \citet{Zahid2014} at \zra{1.4}{1.7} and \citet{Topping2021} at \zest{1.5} as well as showing the KGES sample in the background. All comparison samples metallicities have been derived using the \MT{} calibration. The gas-phase metallicity of the KURVS-CDFS sample agrees with cosmic evolution of the MZR identified by other studies, with higher redshift galaxies having a lower gas-phase metallicity for a given stellar mass compared to \zest{0} galaxies.

In Figure \ref{fig:MS_MZR} we colour the KGES and KURVS-CDFS samples by their \HA{} star-formation rates to highlight the presence of the fundamental mass--metallicity relation (FMR) between stellar mass, gas-phase metallicity and star-formation rate \citep[e.g.][]{Curti2019,Gillman2021,Alaina2021}. Within the KURVS-CDFS sample the FMR is not prevalent due to the small dynamic range of star-formation rates but is visible in the KGES comparison sample.

\col{To assess whether the KURVS-CDFS galaxies lie on the local FMR, we utilise the parameterisation of the plane by \citet{Curti2020} in SDSS galaxies. The plane is defined as 
\begin{equation}\label{eq:FMR}
Z(M,SFR)\,=\,Z_{0}-(\gamma/\beta)\log(1+(M_*/M_0(SFR))^{-\beta})    
\end{equation}
where $\log(M_{0}(SFR))$\,=\,$m_0$+$m_1\log$(SFR). \citet{Curti2020} measured the best-fitting parameters are $Z_0$\,=\,8.779\,$\pm$\,0.005, $m_0$\,=\,10.11\,$\pm$\,0.03, $m_1$\,=\,0.56\,$\pm$\,0.01, $\gamma$\,=\,0.31\,$\pm$\,0.01 and $\beta$\,=\,2.1\,$\pm$\,0.4. 
Using the median stellar mass and star-formation rate of the KURVS-CDFS sample we can use Equation \ref{eq:FMR} to predict the gas-phase metallicity for galaxies on the FMR. At $\log_{10}$(\Mstar[$M_{\odot}$])\,=\,10.1\,$\pm$\,0.1 and \med{SFR$_{\rm H\alpha}$[M$_{\odot}$yr$^{-1}$]}{24}{6}, Equation \ref{eq:FMR}  predicts Z\,=\,8.53\,$\pm$\,0.01 whilst the median KURVS metallicity is Z$_{\rm M13}$\,=\,8.40\,$\pm$\,0.03. This indicates that the KURVS-CDFS galaxies lie $\approx$0.1\,dex below the \zest{0} FMR, however we note this maybe be driven by differences in metallicity indicators \citep[e.g.][]{Kewley2008,Andrews2013, Poetrodjojo2021}}.

\begin{figure*}
    \centering
    \includegraphics[width=\linewidth]{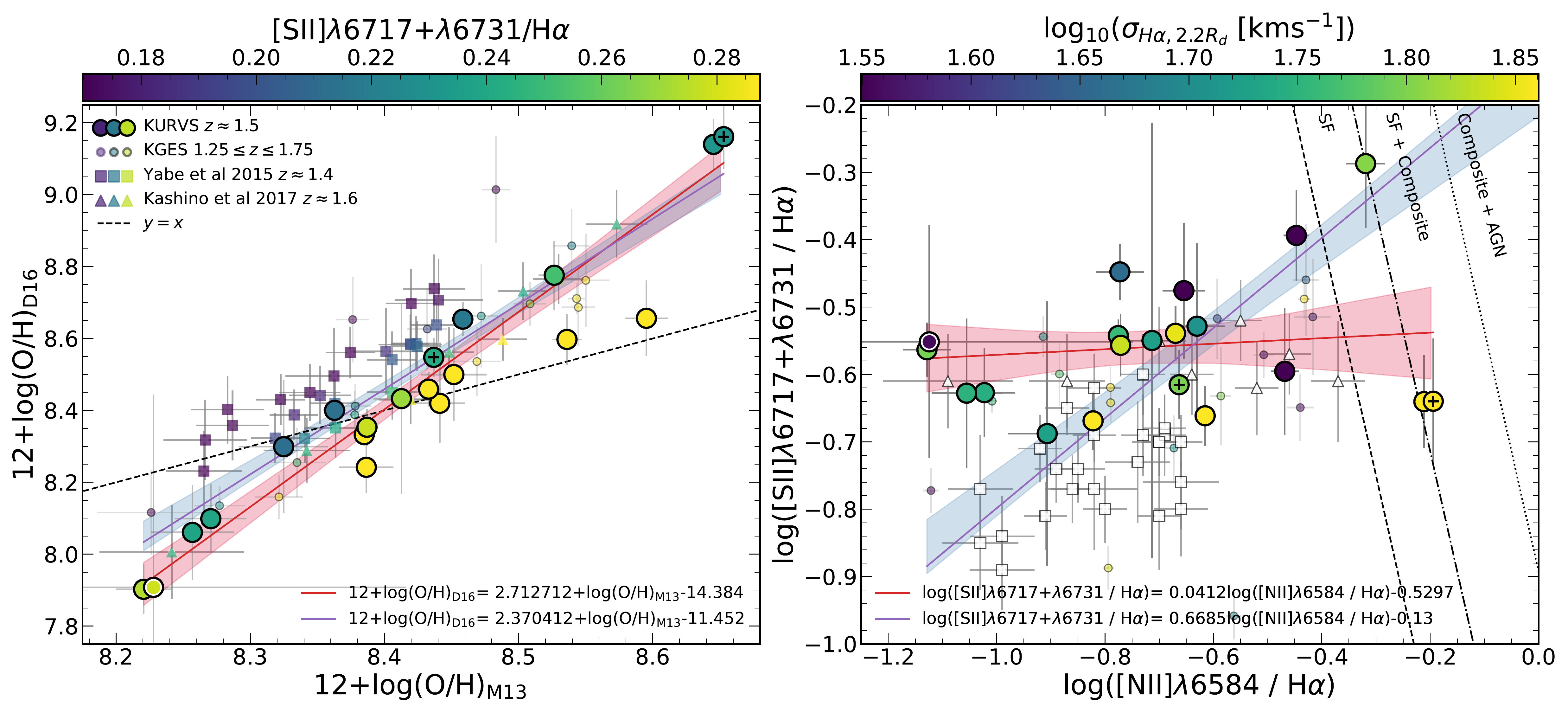}
    \caption{\textbf{Left}: Integrated gas-phase metallicity derived from the \NII, \SII{} and \HA{} strong-line calibration of \citet{Dopita2016} (\DS) as a function of the \citet{Marino2013} (\MT) calibration that utilises the \NH{} ratio, for the star-forming (\textit{circles}) and AGN-candidates (\textit{circles+cross}; Section \ref{Sec:AGN}) in the KURVS-CDFS sample as well as the KGES comparison sample (\textit{dots}). The white outlined circle indicates a $2-\sigma$ limit derived on the \NII{} line for objects with low S/N (see Table \ref{Table:KURVS_metals}). We also show two comparison samples from the \zest{1.6} \citet{Kashino2017} FMOS-COSMOS sample ($triangles$) and the \citet{Yabe2015}  Subarau-FMOS sample ($squares$) at \zest{1.4}. The emission-line ratios for both comparison samples are derived from stacked spectra which we convert to gas-phase Oxygen abundances for the samples using Equations \ref{Eqn:N2} $\&$ \ref{eqn:DP16_OH}. The dashed line indicates a one-to-one relation. We show a fit (and 1-$\sigma$ uncertainty) to the KURVS-CDFS galaxies (red line) as well as the full sample (KURVS-CDFS + comparison samples; blue line). There is good agreement between the two indicators for intermediate gas-phase metallicity galaxies, whilst at high (low) \MT{} metallicity, the \DS{} calibration produces a higher (lower) gas-phase metallicity.  We colour the galaxies by their integrated \SH{} ratio, a strong tracer of the ionization parameter in the interstellar medium. At high metallicity, galaxies with higher \SH{} ratio show closer agreement between the two indicators, whilst at low metallicity galaxies with lower \SH{} ratios show a smaller offset from the one to one relation.
    \textbf{Right}: The \SH{} ratio as a function of the \NH{} ratio for the KURVS-CDFS and comparison samples \col{(white symbols)}. 
    The dashed, dashed-dotted and dotted lines indicate the separation between star-forming, star-forming and composite and composite and AGN galaxies respectively, as derived by \citet{Lopez2010}. The majority of KURVS-CDFS galaxies lie in the star-forming region, with three objects, KURVS 6, 12 and 22 falling in the star-forming composite region. We show linear fits to the KURVS-CDFS sample (red line) and full sample (blue line), with the KURVS-CDFS galaxies indicating no correlation between the two emission line ratios ($r_{\rm s}\,=\,0.17,\,p_{\rm s}\,=\,0.47$) whilst the full sample shows a stronger positive correlation ($r_{\rm s}\,=\,0.71,\,p_{\rm s}\,$<$\,0.001$). We colour the KURVS-CDFS galaxies by their  integrated \HA{} velocity dispersion  (\sig{}).  We identify no correlation between the \NH{} ratio and \sig of the KURVS-CDFS galaxies with a spearman rank coefficients of $r_{\rm s}\,=\,0.18,\,p_{\rm s}\,=\,0.42$. Whilst the \SH{} and \sig{} indicate a negative correlation with $r_{\rm s}\,=\,-0.48,\,p_{\rm s}\,=\,0.03$.}
    \label{fig:integrated_metals}
\end{figure*}

\subsubsection{N2S2\HA{} Index}

The calibrations of \citet{Pettini2004,Marino2013} assume that the ionization parameter of the \HII{} region for which the calibration was derived, is the same as that being analysed. However, the \NH{} index varies strongly with the ionization parameter and hardness of the ionizing radiation field, as well as being dependent on the relation between \NO{} and \OH{}. All three of these properties may vary at high redshift thus making the \NH{} strong-line calibration degenerate with other interstellar medium properties. 

A recently developed strong-line calibration aimed to solve these metallicity degeneracy issues is the N2S2\HA{} calibration. \citet{Dopita2016} (hereafter \DS) defined the new gas-phase metallicity calibration as,
\begin{equation}\label{eqn:DP16_OH}
    12+\log(\rm O/H)_{\rm D16}=8.77+\log(\text{\NII}/\text{\SII})+0.264\log(\text{\NII/H}\alpha)\\   %+0.45(y+0.3)^5
\end{equation}
where \NII/\SII{} is defined as \NII$\lambda$6584/\SII$\lambda$6717+$\lambda$6731.
This calibration has an associated uncertanity of 0.12\,dex and is expected to be less sensitive to the ionization parameter due to the \NS{} emission-line ratio whilst being equally unaffected by extinction due to the short wavelength separation. The \NH{} derived metallicity varies by $\sim$1 dex with ionization parameter \citep{Kewley2019b}. By combining the \NII/\SII{} ratio, that has a stronger dependence on the ionization parameter than \NII/[O{\sc{ii}}] ratio, with the \NH{} ratio, the \DS{} calibration shown in Equation \ref{eqn:DP16_OH} becomes insensitive to the ionization parameter and interstellar medium pressure.

However, the calibration is still strongly dependent on the assumed \NO{} to \OH{} relation, as well as a constant Sulphur to Oxygen ratio, which is calibrated from a sample of local galaxies \citep[e.g.][]{Izotov2006, Maiolino2019}. 
The median gas-phase metallicity derived using the \DS{} calibration for the KURVS-CDFS sample is \med{12+log(O/H)$_{\rm D16}$}{8.43}{0.07}, which is comparable to the median \MT{} metallicity. 

To analyse the conditions of the interstellar medium in the KURVS-CDFS galaxies, in Figure \ref{fig:integrated_metals} we plot the gas-phase metallicity as derived from \DS{} calibration, as a function of the \MT{} derived metallicity. Differences between these two calibrations provide information on the ionization parameter and the hardness of the radiation field. We also show comparison samples from \citet{Yabe2015} at \zest{1.4} from the Subaru FMOS Galaxy Redshift survey  and \citet{Kashino2017} at \zest{1.6} from the FMOS-COSMOS survey, as well as the KGES comparison sample. The emission-line ratios for both literature comparison samples are derived from stacked spectra which we convert to gas-phase Oxygen abundances for the samples using Equations \ref{Eqn:N2} $\&$ \ref{eqn:DP16_OH}. 

The KURVS-CDFS galaxies and comparison samples indicate a comparable relationship between the \DS{} and \MT{} metallicity. There is good agreement between the two indicators for intermediate gas-phase metallicity galaxies, whilst at high (12+log(O/H)$_{\rm M13}>8.5$) and low  (12+log(O/H)$_{\rm M13}<8.3$) \MT{} metallicity, the \DS{} calibration produces a higher (lower) gas-phase metallicity. We perform a linear orthogonal distance regression fit to the KURVS-CDFS galaxies of the form 12+log(O/H)$_{\rm D16}$\,=\,$\alpha$ (12+log(O/H)$_{\rm M13}$) + $\beta$, deriving an $\alpha\,=\,2.71 \pm 0.21$, as indicated by the red line. Whilst the inclusion of the comparison samples results in a slope of  $\alpha\,=\,2.37 \pm 0.14$, \col{as indicated by the blue line}.

\begin{figure}
    \centering
    \includegraphics[width=\linewidth]{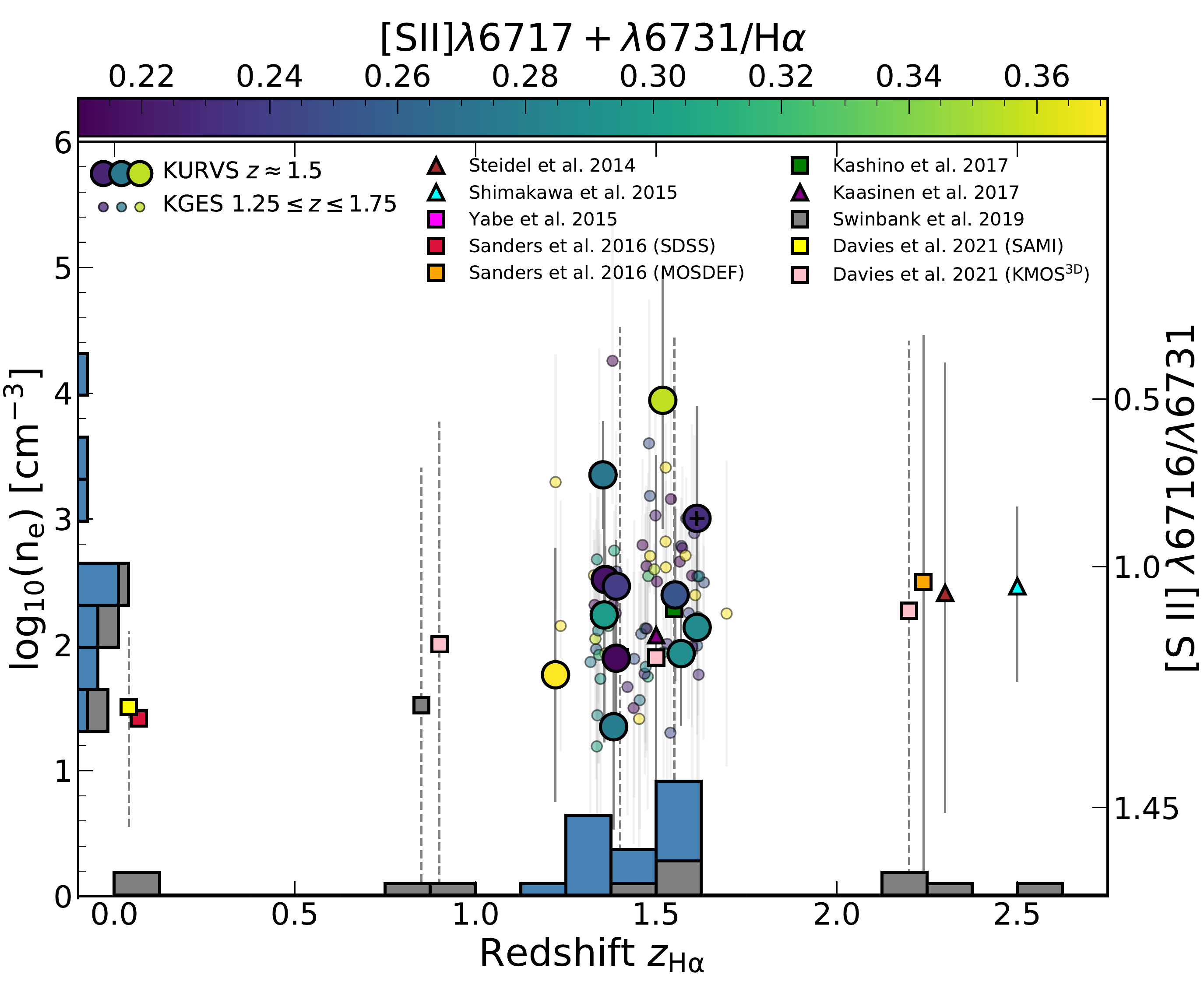}
    \caption{The electron density derived from the \SIIR{} ratio in the 12 KURVS-CDFS sample (\textit{circles}) with 0.45\,$\leq$\,\SIIR$\,\leq$\,1.45, derived following the prescriptions of \citet{Kewley2019a}, as a function of redshift. The circles+cross indicate the candidate AGN in the KURVS-CDFS sample. 
   We include other literature samples of electron density measurements from \OIIR{} (\textit{squares}) and \SIIR{} (\textit{triangles}) ratios measured either from stacks (\textit{dashed errorbars}) or individual galaxies (\textit{solid errorbars}). The uncertainty in the comparison samples indicates the $1-\sigma$ distribution of electron density about the median, in contrast to the KURVS-CDFS errorbars that indicate the uncertainty on the individual \nel{} measurement.  We also show the distribution of electron density from KGES surveys (\textit{dots}) and a histogram  is shown on each axis respectively for the KURVS-CDFS sample (blue) and comparison samples (grey)
   The electron density in the KURVS-CDFS sample agrees with that found by other studies with a median electron density of \med{log$_{10}$(\nel[cm$^{-3}$])}{2.29}{0.25} at a redshift of \med{$z$}{1.39}{0.07}, which is higher than local galaxies. The KURVS-CDFS (and KGES) galaxies are coloured by their integrated \SH{} ratio, a proxy for the ionization parameter of the interstellar medium, however no strong trend is identified.}
    \label{fig:ne_evolution}
\end{figure}

In Figure \ref{fig:integrated_metals} we colour the KURVS-CDFS galaxies, and comparison samples, by their \SH{} ratio. The \SH{} ratio is known to be a poor tracer of gas-phase metallicity due to its strong dependence on the ionization parameter \citep[e.g.][]{Maiolino2019}. At higher metallicities, the galaxies with higher \SH{} ratios show closer agreement between the two indicators, whilst the low \SH{} ratio galaxies exhibit the largest discrepancy. At the lowest metallicities the lower \SH{} ratio galaxies have a smaller offset from the one to one relation. On average the KURVS-CDFS galaxies have higher \SH{} ratios (harder ionizing spectrum) with a median \med{\SH}{0.27}{0.02} compared to \med{\SH}{0.20}{0.01} for the comparison samples. The \MT{} calibration assumes fixed N/O - O/H relation whereas the \DS{} calibration ratio accounts for variation in this relation which may result in the offset between the two metallicity indicators. \col{Furthermore, \DS{} is based on grids from the  {\sc{mappings}} photoionisation models, in contrast to the \MT{} calibration which is based on local HII regions. Thus the dynamic range in metallicity is different and hence the two calibrations differ the most in the low- and high metallicity regimes.}

\subsubsection{Emission-Line Diagnostics}

To further understand the properties of the interstellar medium in KURVS-CDFS galaxies, we can compare different integrated  emission-line ratios. High-redshift galaxies are offset to higher emission-line ratios compared to local main-sequence galaxies however the origin of this offset is disputed with indications of harder ionizing spectrum, higher ionization parameter, and/or variable N/O abundance ratio \cite[e.g.][]{Kewley2013a,Steidel2014,Strom2017,Topping2020,Izotov2021,Runco2021} whilst other studies suggest the offset is purely an observational selection effect \citep[e.g.][]{Garg2022}.

Sulfur is one of the $\alpha$ elements, which include e.g. O, Ne, Si, produced primarily through nucleosynthesis in massive stars and supplied to the interstellar medium through type-II supernovae. In contrast, Nitrogen is generated through both the primary process and a secondary process where $^{12}$C and $^{16}$O initially contained in stars are converted into $^{14}$N via the CNO cycle. Therefore the \NII{}/\SII{} ratio is sensitive to the total chemical abundance, particularly in a regime where secondary nitrogen is predominant, while this ratio is almost constant if most nitrogen has a primary origin \citep[e.g][]{Strom2021}. The \SII{}/\NII{} ratio has only a small effect by dust as the emission lines are separated by $\sim$140\AA.\\

One useful diagnostic that also provides information on the origin of the ionizing radiation field  and can distinguish between star-forming galaxies and active galactic nuclei (AGNs) is the correlation between \SH{} ratio and the \NH{} emission-line ratio \citep[e.g.][]{Sabbadin1977}.

In the right-panel of Figure \ref{fig:integrated_metals} we show the galaxy integrated \SH{} ratio as a function of \NH{} ratio, coloured by the galaxies velocity dispersion {\sig} measured at 2.2R$_{\rm d}$, as derived in \AP, where R$_{\rm d}$ the stellar continuum disc scale length of the galaxy. We also show the comparison samples from \citet{Yabe2015} at \zest{1.4} and \citet{Kashino2017} at \zest{1.6} as well as the KGES comparison sample. We perform a linear orthogonal distance regression fit to the data of the form \SH{}=\,$\alpha$\,(\NH)\,+$\beta$. The KURVS-CDFS galaxies indicate no strong correlation between the emission-line ratios with an $\alpha\,=\,0.048 \pm 0.088$, as indicated by the red line and spearman rank correlation ($r_{\rm s}, p_{\rm s}$) coefficents of $r_{\rm s}\,=\,0.17,\,p_{\rm s}\,=\,0.47$.  The inclusion of the full sample (KURVS-CDFS + literature samples) leads to a positive correlation with $\alpha\,=\,0.67 \pm 0.045$ (blue line) and spearman rank coefficents of of $r_{\rm s}\,=\,0.71,\,p_{\rm s}\,=\,0.0$.

Three KURVS-CDFS galaxies, KURVS 6, 12 and 22 lie at the boundary of the star-forming to star-forming and composite. KURVS 12 was identified in Section \ref{Sec:AGN} as a candidate AGN, whilst KURVS 22 has an X-ray counterpart, but with a luminosity below the \citet{Luo2017} AGN threshold. KURVS 6 indicates no AGN spectral features, however the rest-frame optical morphology of this galaxy shows a clear bulge component which may result in higher emission-line ratios \citep[e.g.][]{Mendez2019,Pak2021}. 

\citet{Law2021} identify strong positive correlations between the gas-phase velocity dispersion  and  emission-line ratios such as \SH, \NH{} in 9149 local ($z\,=\,0.04$) galaxies in the MANGA survey with a stellar mass range of $\rm \log_{10}(M_*[M_{\odot}])$\,=\,9\,--\,11. They suggest the origin of the correlation is driven by the variation in ionizing source across the emission-line ratio parameter space. From Figure \ref{fig:integrated_metals}, we identify no correlation between the \NH{} ratio and \sig of the KURVS-CDFS galaxies with a spearman rank coefficients of $r_{\rm s}\,=\,0.18,\,p_{\rm s}\,=\,0.42$. Whilst the \SH{} and \sig{} indicate a negative correlation with $r_{\rm s}\,=\,-0.48,\,p_{\rm s}\,=\,0.03$ in contrast to \citet{Law2021} at \zest{0}. We note however we are comparing integrated emission-line ratios and velocity disperions as opposed to the spaxel-wise ratios compared in \citet{Law2021}.

\begin{figure*}
    \centering
    \includegraphics[width=\linewidth]{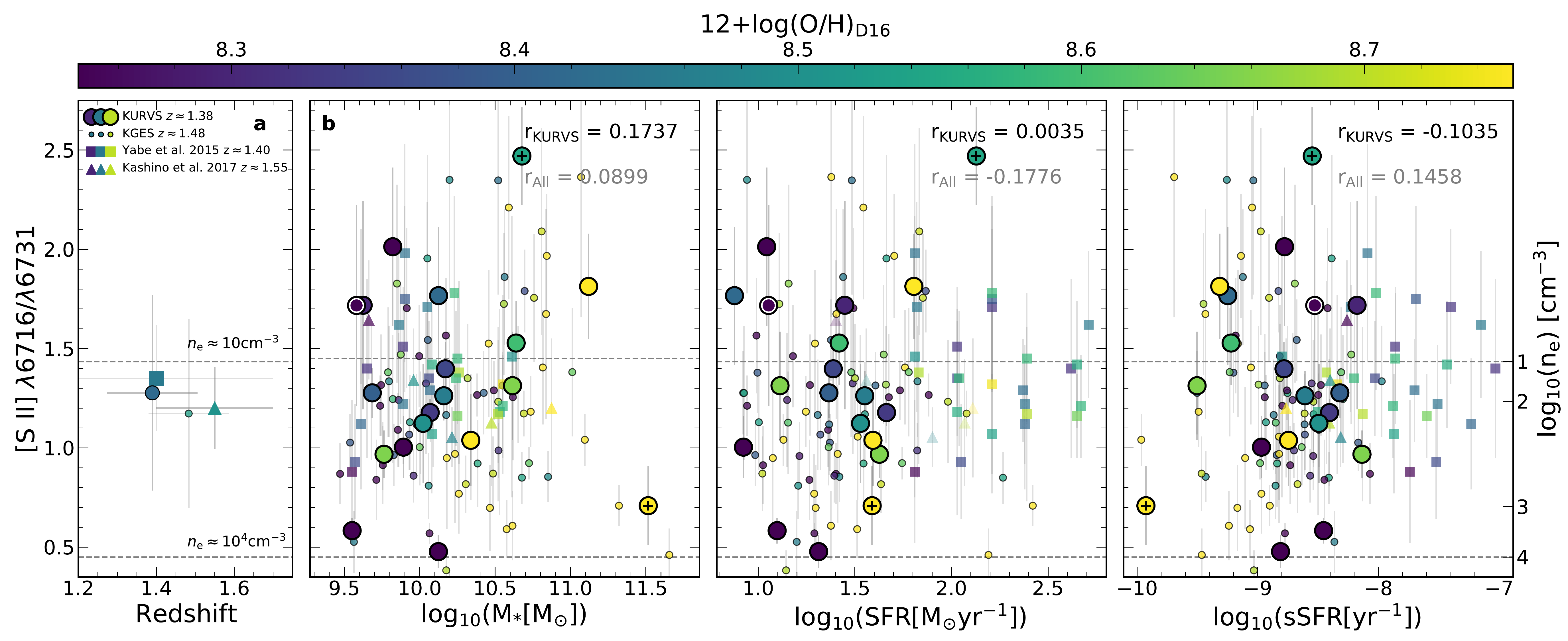}
    \caption{The \SIIR{} ratio for the KURVS-CDFS sample, and literature samples, as a function of main-sequence galaxy properties. We first show the median \SIIR{} ratio as a function of median spectroscopic redshift, with errors indicating the $1-\sigma$ distribution (\textbf{a}). We then show the \SIIR{} for individual KURVS-CDFS galaxies as a function of stellar mass (\textbf{b}), \HA{} star-formation rate (\textbf{c}) and \HA{} specific star-formation rate (\textbf{d}). The white outlined circle indicates a $2-\sigma$ limit derived on the \NII{} line for object with low S/N (see Table \ref{Table:KURVS_metals}). The theoretical limits of the electron density model from \citet{Kewley2019a}; \SIIR{}\,=\,0.45 and \SIIR\,=\,1.45, are indicated by the dashed black lines. We find no trend between all three quantities as indicated by the black spearman rank correlation coefficients in the top corner of each panel.  We also show the \SIIR{} for individual KGES galaxies, and the results from stellar mass stacked spectra in the \citet{Kashino2017} \zest{1.6} sample and the \citet{Yabe2015} sample at \zest{1.4}.  The inclusion of these samples also results in no correlations between the  \SIIR{} ratio and galaxy properties, as shown by the grey spearman rank correlation coefficients. The galaxies are coloured by their gas-phase metallicity derived following the \DS{} calibration however no correlation between \SIIR{} and gas-phase metallicity is identified.}
    \label{fig:ne_prop}
\end{figure*}

\subsection{Galaxy Integrated Electron Density}\label{Sec:edensity}

As well as variations in the ionising source and the hardness of the radiation field, the density of electrons (\nel{}) in the interstellar medium of a galaxy can affect the inferred metallicity from emission-line ratios. At high-redshift an increased ionization parameter may be responsible for the higher emission-line ratios observed in main-sequence galaxies, in addition to the increased star-formation efficiency at earlier times. A possible driver of a higher ionization parameter is an elevated electron density \citep[e.g.][]{Brinchmann2008,Kashino2017,Kassinen2017,Kewley2019a}.

Local galaxies have electron densities in the range of log$_{10}$(\nel[cm$^{-3}$])\,$\approx$\,1.5\,--\,2, which corresponds to variation in the ionization parameter of $\approx$0.06\,dex. Recent high-redshift studies suggest an electron density of log$_{10}$(\nel[cm$^{-3}$])\,$\approx$\,3 corresponding to a $\approx$0.5\,dex increase in the ionization  parameter \citep[e.g.][]{Sanders2016,Kewley2019a,Davies2021}. 
In this section we quantify the electron density for individual galaxies in the KURVS-CDFS sample and analyse the correlations with the galaxy's main sequence properties.

The electron density can be estimated from the ratio of collisionally-excited metal lines (e.g. \OIIR{}, \SIIR{}) \citep{Osterbrock1974}. These emission-line ratios weakly depend on electron temperature and are largely independent of metallicity. If we assume a constant nebula temperature of T\,=\,10$^4$\,K, the interstellar medium pressure is then directly proportional to the electron density. Doing so allows us to convert the \SIIR{} ratio to an electron density following the ionization models of \citet{Kewley2019a}. The \SIIR{} line ratio is density sensitive in the range log$_{10}$(\nel[cm$^{-3}$])\,$\approx$\,1.6\,--\,3.7 corresponding to \SIIR{}\,=\,1.45\,--\,0.45 respectively.

We note however this is a simplification, as \HII{} regions contain temperature gradients and complex ionization structures \citep[e.g.][]{Wang2004,Dopita2006b,Kewley2019a}.
In addition the \SII{} lines cannot probe high densities because collisional de-excitation dominates above 10$^4$cm$^{-3}$ where the ratio saturates at $\sim$0.45, its asymptotic value \citep[e.g.][]{Davies2020}. 
To measure the integrated electron density for each galaxy, we take the \SIIR{} ratio as reported in Table \ref{Table:KURVS_metals} and apply the conversion to electron density. We detect the \SII{} lines in 19 KURVS-CDFS galaxies with \SII{} S/N$\geq$2.
The \citet{Kewley2019a} model places theoretical constraints on the  emission-line ratios, requiring them to be between 0.45\,$\leq$\,\SIIR$\,\leq$\,1.45. We therefore exclude the 10 galaxies that lie outside of these theoretical limits (8 with \SIIR$>$1.45, 2 with \SIIR$<$0.45\footnote{The \SII{} lines in KURVS 2 and 19 are heavily affected by OH skylines}). The median \nel{} of the remaining 12 KURVS-CDFS galaxies is \med{log$_{10}$(\nel[cm$^{-3}$])}{2.29}{0.25} at a median redshift of \med{$z$}{1.39}{0.07} 

In Figure \ref{fig:ne_evolution} we plot the electron density of the KURVS-CDFS galaxies as a function of their spectroscopic redshift as well the electron density measurements from individual galaxies in the KGES comparison sample. In addition to the KMOS samples we also include a number of comparison samples derived from either the \OIIR{}  and \SIIR{} for both stacked and individual spectra, calibrated to the \citet{Kewley2019a} model by \citet{Davies2021}.  The electron density within the KURVS-CDFS galaxies agrees with comparison studies, with a higher electron density at \zest{1.5} than in the local Universe. The KMOS samples are coloured by their integrated \SH{}, a proxy for the ionization parameter, however no distinct trend with redshift or electron density is identified. 

As well as a cosmic evolution in the electron density of the interstellar medium, studies have shown that the electron density may correlate with the galaxies main-sequence properties  \citep[e.g.][]{Shimakawa2015,Kassinen2017,Davies2021} whilst other studies conflict these results, finding no such correlations \citep[e.g.][]{Yabe2015,Onodera2016,Sanders2016}. In Figure \ref{fig:ne_prop} we first plot the median \SIIR{} ratio of the KURVS-CDFS galaxies, and comparison samples, as a function of the median redshift, coloured by their gas-phase metalliciy derived using the \DS{} calibration. On average the KURVS-CDFS galaxies have similar electron density and gas-phase metallicity to the other comparison samples at this epoch (see Table \ref{Table:ne_prop}).

We then correlate the individual galaxies \SIIR{} ratios with their  their stellar mass, \HA{} star-formation rate and \HA{} specific star-formation rate. The galaxies are coloured by their integrated gas-phase metallicity following the \DS{} calibration. We include comparison samples  from \citet{Kashino2017} FMOS-COSMOS and \citet{Yabe2015} the Subaru FMOS Galaxy Redshift Survey as well as the KGES comparison sample. The \SIIR{} is plotted as opposed to the \nel{} to remove the theoretical constrains of the \citet{Kewley2019a} model, allowing more galaxies to be included in the sample. We note the electron density shown for the KMOS galaxies represents individual galaxy measurements, whilst both comparison samples are derived from spectra stacked in stellar mass bins.

For both the KURVS-CDFS sample and full sample (KURVS-CDFS + literature samples), we identify no significant correlations in all three panels as indicated by the black and grey spearman rank correlations shown in each panel respectively. We further identify no connection with between the integrated gas-phase metallicity and \SIIR{} ratio of the galaxies. However we note the KURVS sample does not cover a large range in stellar mass and star-formation rate, and thus may lead to a lack of correlation.

\begin{table}
\caption{Median redshift, \SII{} ratio and gas-phase metallicity of KMOS and comparison samples.}
\label{Table:ne_prop}
\begin{tabular}{|llll|}
\hline
Survey & $z_{\rm spec, H\alpha}$ &  \SII $\frac{\lambda 6717}{\lambda 6731}$ & 12+log(O/H)$_{\rm D16}$ \\
\hline
KURVS-CDFS & 1.39 $\pm$ 0.11 & 1.27 $\pm$ 0.80 & 8.43 $\pm$ 0.33 \\
KGES & 1.48 $\pm$ 0.10 & 1.17 $\pm$ 0.48 & 8.54 $\pm$ 0.28 \\
\citet{Yabe2015}& 1.44 & 1.35 $\pm$ 0.27 & 8.45 $\pm$ 0.13 \\
\citet{Kashino2017}& 1.55 & 1.20 $\pm$ 0.21 & 8.51 $\pm$ 0.25 \\
\hline
\end{tabular}
\end{table}

\subsection{Resolved Gas-Phase Metallicity}

To constrain the role of the baryon cycle within high-redshift star-forming galaxies it is crucial to spatially resolve the properties of the interstellar medium. The target selection of the KURVS-CDFS sample means all galaxies have deep ancillary multi-wavelength data, in addition to the unprecedented depth of the integral field KMOS observations, examples of which are  shown in Figure \ref{fig:HST_resolved}. In this section we exploit this data to quantify the spatially resolved properties of the interstellar medium at \zest{1.5}.

\subsubsection{Resolved Fundamental Metallicity Relation} \label{Sec:resolved_FMR}

The local galaxy population is well defined by a number of well studied scaling relations that interconnect the fundamental properties of galaxies \citep[e.g.][]{Fall2013,Fall2018,Curti2020}. The fundamental metallicity relation (FMR) between galaxy stellar mass, gas-phase metallicity and star-formation rate prescribes the relation between the chemical abundance of the interstellar medium gas, ongoing star formation and stellar mass build up.
The spatially resolved nature of the fundamental metallicity relation is less well defined especially at high redshift.

To resolve the FMR within the KURVS-CDFS sample, we require spatially resolved stellar mass, metallicity and star-formation maps. The metallicity and star-formation maps can be derived from the integral field KMOS data, whilst we derive the stellar mass maps from high-resolution CANDELS imaging of the KURVS-CDFS sample. Using the deep multi-wavelength broadband \textit{HST} imaging from 0.43\,--\,1.6$\mu$m, Dudzevi\v{c}i\={u}t\.{e} in. prep. perform pixel-to-pixel SED fitting. \col{To do so we re-binned all of the \textit{HST} data to 0.18 arcseconds, to approximately match the PSF FWHM of the longest wavelength ($H$-band) data.  From this matched-resolution imaging, the \sex{} software was used to perform pixel photometry and stellar mass maps were constructed using  using the \magphys{} \citep{daCunha2015} SED fitting code (Dudzeviciute et al in prep)}. \citet{Dudzevivciute2019} confirmed that the spatially resolved stellar mass values of galaxies in the CANDELS field are consistent with integrated values.

\begin{figure*}
    \centering
    \includegraphics[width=\linewidth]{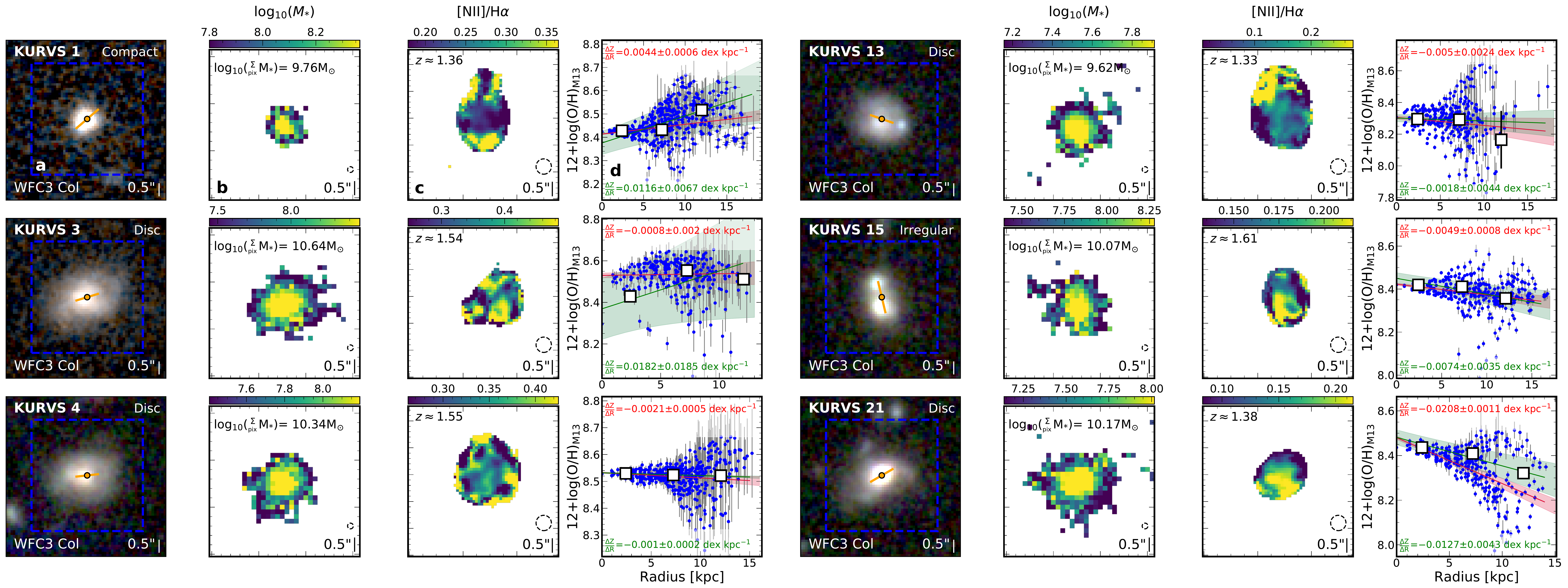}
    \caption{Examples of the \textit{HST} multi-band imaging and spatially resolved KMOS data for the KURVS-CDFS galaxies. For each galaxy we show a WFC3 three-colour image composed of F105W, F125W and F160W images \col{with the KMOS field of view (blue square) overlaid }(\textbf{a}). The semi-major axis of the galaxy (orange line) and stellar continuum centre (black filled circle) are indicated, as well as the morphological classification from \citet{Huertas-Company2015}. The spatially resolved stellar mass map  (\textbf{b}) and \NH{} map  (\textbf{c}) are shown with the FWHM of the PSF in \textit{HST} and KMOS  observations (black dashed circle) respectively. In the final panel (\textbf{d}), we display the pixel-based gas-phase metallicity (blue points) as a function of radius and linear fit (crimson line) with reported slope and uncertainty. \col{We also show the annuli-based metallicity (black squares) and corresponding annuli-based gradient (green line) with slope and uncertainty}. The examples indicate the range of morphologies and metallicity profiles in addition to the KMOS data quality.}
    \label{fig:HST_resolved}
\end{figure*}

To derive the gas-phase metallicity maps for each galaxy we convert the \NH{} ratio derived in each pixel (see Figure \ref{fig:HST_resolved}), following the adaptive binning emission-line fitting procedure (see \AP),  to a gas-phase Oxygen abundance using the \MT{} calibration. We opt for the \MT{} calibration as we do not have sufficient S/N in the \SII{} lines to spatially resolve them. The \HA{} star-formation rate map is derived by converting the observed \HA{} flux in each pixel to a extinction corrected \HA{} star-formation rate, following the same procedures as in Section \ref{Sec:MSFG} but using each pixels \magphys{} derived $A_{V}$ to correct for dust-extinction.

In order to correlate the stellar mass in each pixel with the ionised gas properties (e.g. metallicity, star-formation rate) of the galaxies, we need to match the \textit{HST} stellar mass map pixel scale and PSF size with that of the KMOS data. We re-bin the KMOS data, that is sampled at 0.10 arcseconds per pixel, to be 0.18 arcsecond per pixel. In doing so the flux in the KMOS data is conserved. To match the PSF size, we assume the PSFs are well modelled by a Gaussian and smooth the \textit{HST} stellar mass maps with a Gaussian kernel of width,
\begin{equation}
    \theta=(\theta_{\rm KMOS}^2-\theta_{HST}^2)^{0.5}
\end{equation}
where $\theta_{\rm KMOS}=0.58$ arcseconds and $\theta_{HST}=0.22$ arcsecond. The KMOS PSF is derived from the standard star frames in the KMOS observations.
Re-binning the KMOS data to a larger pixel scale results in the loss of pixels at the edges due to contamination with lower S/N pixels. We remove these contaminated pixels as well pixels from the stellar mass map with poorly constrained stellar masses due to uncertainties in the SED fitting.
 
With the \textit{HST} stellar mass maps of the galaxies on the same scale as the kinematic KMOS data, \col{we align the \textit{HST} data with the KMOS data using deep \textit{H}-band continuum images. Having aligned and re-scaled the data to a homogeneous resolution,} we can now analyse the spatially resolved correlations in the KURVS-CDFS sample. \col{In Figure \ref{fig:HST_resolved} we show examples of the \textit{HST} multi-band imaging and spatially resolved KMOS data for the KURVS-CDFS galaxies. For each galaxy we show a WFC3 three-colour image composed of F105W, F125W and F160W images with the KMOS field of view (blue square) overlaid in panel a. The semi-major axis of the galaxy (orange line) and stellar continuum centre (black filled circle) are indicated, as well as the morphological classification from \citet{Huertas-Company2015}. In panels b and c we show the spatially resolved stellar mass map and \NH{} map with the FWHM of the PSF in \textit{HST} and KMOS  observations (black dashed circle) respectively. We display the pixel-based gas-phase metallicity (blue points) as a function of radius and linear fit (crimson line) with reported slope and uncertainty in the final panel d. We also show the annuli-based metallicity (black squares) and corresponding annuli-based gradient (green line) with slope and uncertainty. The full KURVS-CDFS sample is shown in Appendix \ref{App:Resolved_Properties}}

To convert the stellar mass and star-formation rate maps to a stellar mass surface density and star-formation rate surface density maps, we divide each pixels stellar mass (or star-formation rate) by the physical area of each pixel.

\begin{figure*}
\centering
\includegraphics[width=\linewidth]{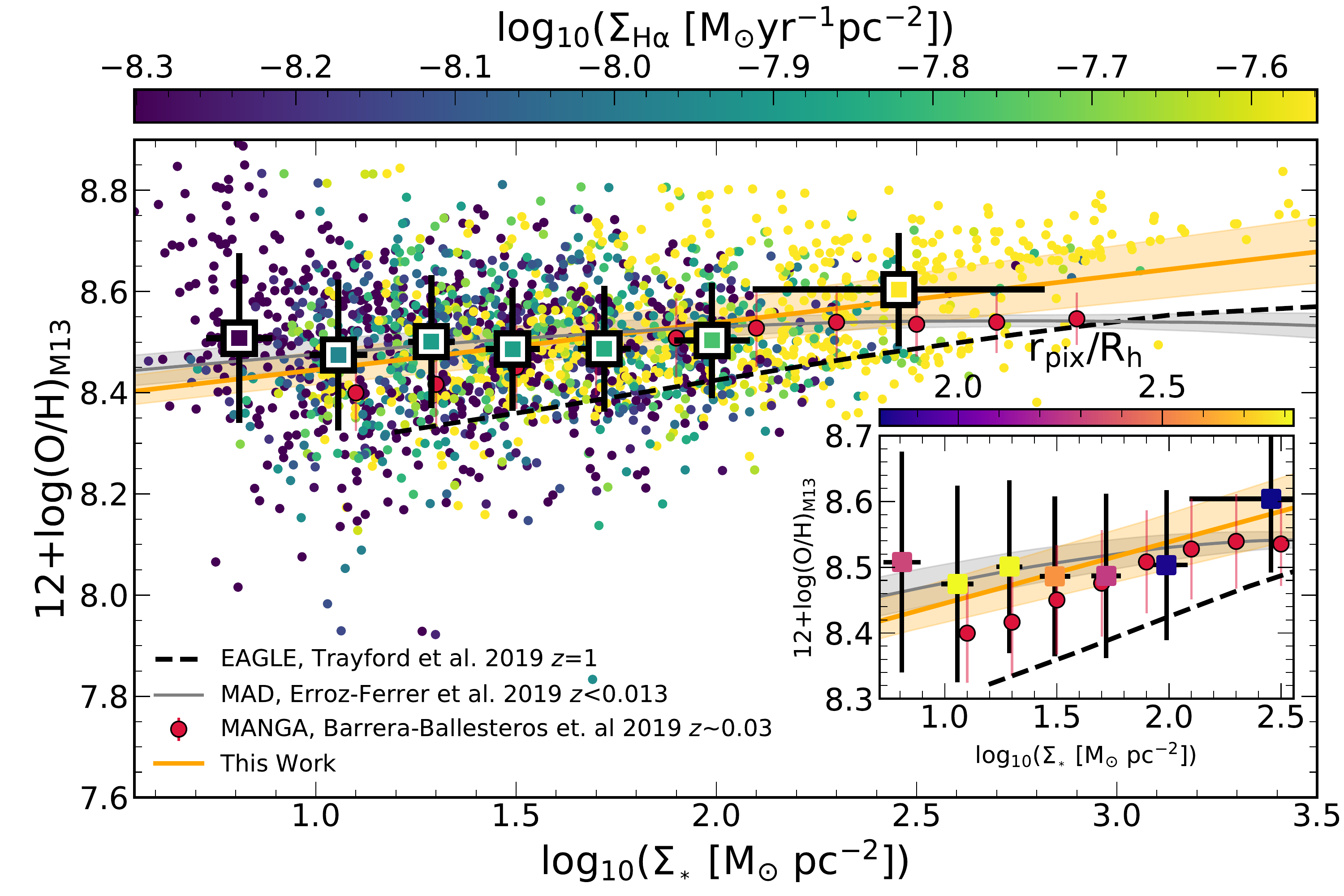}
\caption{The spatially resolved fundamental metallicity relation (rFMR) for the KURVS-CDFS sample. We show the pixel based metallicity (12+log(O/H)), derived following the \MT{} \NH{} calibration, as a function of the stellar mass surface density (log$_{10}$($\Sigma_{_*}$ [M$_{\odot}$ pc$^{-2}$])). The data is coloured by the  pixel-to-pixel \HA{} derived star-formation rate surface density (log$_{10}$($\Sigma_{\rm H\alpha}$ [M$_{\odot}$yr$^{-1}$pc$^{-2}$])).
We perform a orthogonal distance regression fit to the relation (orange line) identifying a slope of $\alpha\,=\,0.09\,\pm\,0.01$ and $\beta\,=\,8.35\,\pm\,0.01$, as well as showing the running median (and standard deviation) in X with 300 pixels per bin (black squares and errorbars). Each bin is coloured by the median star-formation rate surface density in the bin. 
The relations from the MUSE Atlas of Disks survey at $z<0.013$ \citep{Erroz-Ferrer2019} (grey solid line and shaded region), the MANGA survey from \citet{Barrera-Ballesteros2018} at \zest{0.03} (red circles)  and for the {\sc{eagle}} hydrodynamical simulation at $z=1$ from \citet{Trayford2019} (black dashed line) are shown for comparison.  We convert metallicity of \citet{Erroz-Ferrer2019} sample to the \MT{} calibration using the conversion of \citet{Scudder2021}.
The inset panel shows the running median and comparison samples, where the running medians are coloured by the median r$_{\rm pix}$/\Rh{} in each bin.  
The derived relation for the KURVS-CDFS galaxies indicates higher stellar-mass surface density regions having higher gas-phase metallicity and is comparable to the low-redshift comparison samples with on average 0.05\,$\pm$\,0.01\,dex higher metallicity for a given stellar mass surface density. The higher star-formation rate surface density pixels have a lower gas-phase metallicity for a given stellar mass surface density, indicating the presence of the FMR. There is an indication of a correlation between the radial location of the pixels and their metallicity in the inset panel, with highest metallicity located in the inner regions of the galaxies. 
}
\label{Figure:resolvedMZR}
\end{figure*}

\begin{figure*}
    \centering
    \includegraphics[width=\linewidth]{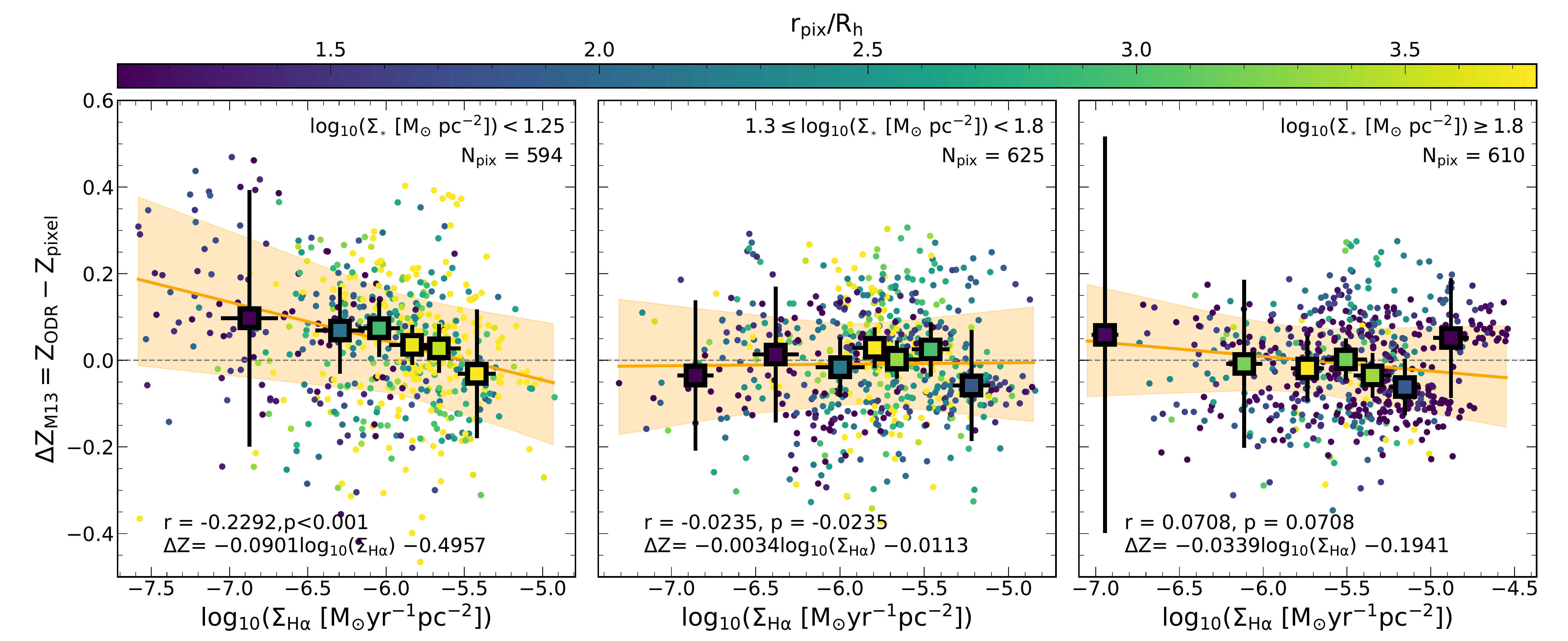}
    \caption{Residuals of the orthogonal distance regression fit to the rFMR (Figure \ref{Figure:resolvedMZR}) as a function of the star formation rate surface density \coln{split into three stellar mass surface density bins}. We show a running median and standard deviation (black squares) in star-formation rate surface density with 100 points per bin \coln{in each stellar mass surface density bin}. We colour the points by their normalised pixel radius ($\rm r_{pix}/R_h$) and show the median value in each median bin, indicating the radial variation in star formation rate surface density.
    We also perform a orthogonal distance regression fit to the relation identifying a slope of \coln{$\alpha\,=\,-0.0901\,\pm\,0.025$, $0.003\,\pm\,0.028$, $0.034\,\pm\,0.024$, and $\beta\,=\,-0.498\,\pm\,0.148$, $-0.011\,\pm\,0.170$, $0.194\,\pm\,0.137$ for each panel respectively.}
    There is a weak negative correlation at high significance \coln{in the lowest stellar mass surface density bin with $r\,=\,-0.229,p\,<\,0.001$, whereby higher star formation rate surface density pixels have a lower gas-phase metallicity for a given stellar mass surface density. This anti-correlation between star formation rate surface density and gas-phase metallicity residual is not present in the medium or higher stellar mass surface density bins. }}
    \label{fig:rFMR_resid}
\end{figure*}

In  Figure \ref{Figure:resolvedMZR} we plot the gas-phase metallicity in each pixel as a function of the stellar mass surface density of the pixel coloured the pixels star-formation rate surface density. A running median \col{(and standard deviation)} in stellar mass surface density is shown by the black squares, where each bin contains 300 pixels. Each bin is coloured by the median star-formation rate surface density in that bin. We also perform a linear orthogonal distance regression fit to the resolved FMR of the form $\rm 12+log(O/H)\,=\,\alpha\,log_{10}(\Sigma_*)+\beta$, deriving $\alpha\,=\,0.09\,\pm\,0.01$ and $\beta\,=\,8.35\,\pm\,0.01$. \col{We note the choice of pixel scale does not influence the derived scaling relations with a consistent relation derived when using a pixel scale comparable to the KMOS PSF.}

In addition we show the relations for the MUSE Atlas of Disks survey at $z<0.013$ from \citet{Erroz-Ferrer2019} (grey solid line and shaded region), the  MANGA survey from \citet{Barrera-Ballesteros2018} at \zest{0.03} (red circles) and from the {\sc{eagle}} hydrodynamical simulation at $z=1$ from \citet{Trayford2019} (black dashed line).  We convert metallicity of \citet{Erroz-Ferrer2019} observational sample to the \MT{} calibration using the conversion of \citet{Scudder2021}. 

The derived relation for the KURVS-CDFS galaxies is in close agreement with the comparison samples, with higher stellar-mass surface density regions having higher gas-phase metallicity. 
We see no strong redshift evolution in the resolved FMR compared to the local Universe, with the low-redshift sample of \citet{Barrera-Ballesteros2018} having $\approx$\,0.05\,dex lower  gas-phase metallicity for a given stellar mass surface density.  \citet{Trayford2019} note the normalisation of the relation for \eagle{} is not necessarily accurate and has been adjusted to match observational metallicity calibrations, it is the shape of the relation, that indicates a weaker slope than our relation, that is important. The minimal evolution of the resolved FMR was also identified in \citet{Patrico2019} for three lensed galaxies at \zra{0.6}{1.0}. \col{However, the evolution of rMZR with redshift is in contrast to the evolution in the integrated MZR (Figure \ref{fig:MS_MZR}) for the KURVS-CDFS sample, where for a given stellar mass, high-redshift galaxies have a lower metallicity. \citet{Trayford2019} identified a strong evolution of 0.4\,dex in metallicity at a stellar mass density of 10$^{2}$ $ \rm M_*pc^{-2}$ between $z=0.1$ and $z=2$. For the same range of stellar mass surface density, we identify an average offset to higher metallicity of $\approx$\,0.12\,dex than that predicted by \citet{Trayford2019} at $z=1$ which may be driven by the choice of metallicity calibration.}

For a given stellar mass surface density in Figure \ref{Figure:resolvedMZR}, the higher star-formation rate surface density pixels \col{appear to} have a lower gas-phase metallicity. \col{To investigate this correlation further, in Figure \ref{fig:rFMR_resid} we correlate the residuals to the orthogonal distance regression fit in Figure \ref{Figure:resolvedMZR} with the star formation rate surface density of the pixels} \coln{in different stellar mass surface density bins.}

We perform a orthogonal distance regression fit to the relation identifying a slope of \coln{$\alpha\,=\,-0.0901\,\pm\,0.025$, $0.003\,\pm\,0.028$, $0.034\,\pm\,0.024$, and $\beta\,=\,-0.498\,\pm\,0.148$, $-0.011\,\pm\,0.170$, $0.194\,\pm\,0.137$ for the low, medium and high stellar mass surface density bins respectively}.  There is a weak negative correlation at high significance between the offset from parametric fit and the pixels star formation rate surface density \coln{in the lowest stellar mass surface density bin with $r\,=\,-0.229,\,p\,<\,0.001$, whereby higher star formation rate surface density pixels have a lower gas-phase metallicity for a given stellar mass surface density. This anti-correlation between star formation rate surface density and gas-phase metallicity residual is not present in the medium or higher stellar mass surface density bins, reflecting the predicted flattening in the rFMR. This is in agreement with that observed at low redshift \citep[e.g.][]{Cresci2019} and indicates the presence of the rFMR in the KURVS-CDFS sample, as shown in Figure \ref{Figure:resolvedMZR}.} We colour the points in Figure \ref{fig:rFMR_resid}, and running median,  by their radial location in each galaxy (normalised by the galaxies stellar continuum half-light radii (\Rh)), indicating the radial variation in star formation rate surface density.

\col{Furthermore,} the inset panel in Figure \ref{Figure:resolvedMZR} shows the running median of the rMZR coloured by the median normalised radius. There is an indication of a correlation between the radial location of the pixels and their metallicity, with highest metallicity located in the inner regions of the galaxies. This maybe due to the `inside-out' secular evolution of the galaxies \citep[e.g.][]{Sanchez2014,Schonrich2017,Zewdie2020, Wang2021}. However, this is averaged over the whole sample so in order to understand this relation further we now analyse the gas-phase metallicity gradients in individual galaxies. We note however this may also be driven by shocks and AGN contamination in the central regions.

\subsubsection{Metallicity Gradients}\label{Sec:profiles}

A negative abundance gradient (metallicity decreasing with radius) has been commonly observed in low-redshift main-sequence star-forming galaxies \citep[e.g.][]{Belfiore2017,Sharda2021,Lutz2021} and is theorised to be a clear signature of the `inside-out' secular evolution of galaxies. The abundance gradients in high-redshift galaxies, however, are less well constrained with a studies reporting a range of abundance gradients \citep[e.g.][]{Cresci2010,Jones2013,Wuyts2016,Curti2019}.

The challenge at high-redshift is the lower S/N and surface brightness of the galaxies, making it difficult to accurately measure emission-line ratios. To achieve the required S/N, one common technique is to stack the spectra from integral field observations inside an annulus. The annuli are spaced apart by the PSF of the observations and aligned to galaxies kinematic position angle and morphological axis ratio \citep[e.g.][]{Wuyts2016,Curti2020,Gillman2021}. We can then measure the emission-line ratios from the stacked spectra in each annulus, giving two or three measurements of the gas-phase metallicity as a function of radius for each galaxy.

\begin{figure*}
    \centering
    \includegraphics[width=\linewidth]{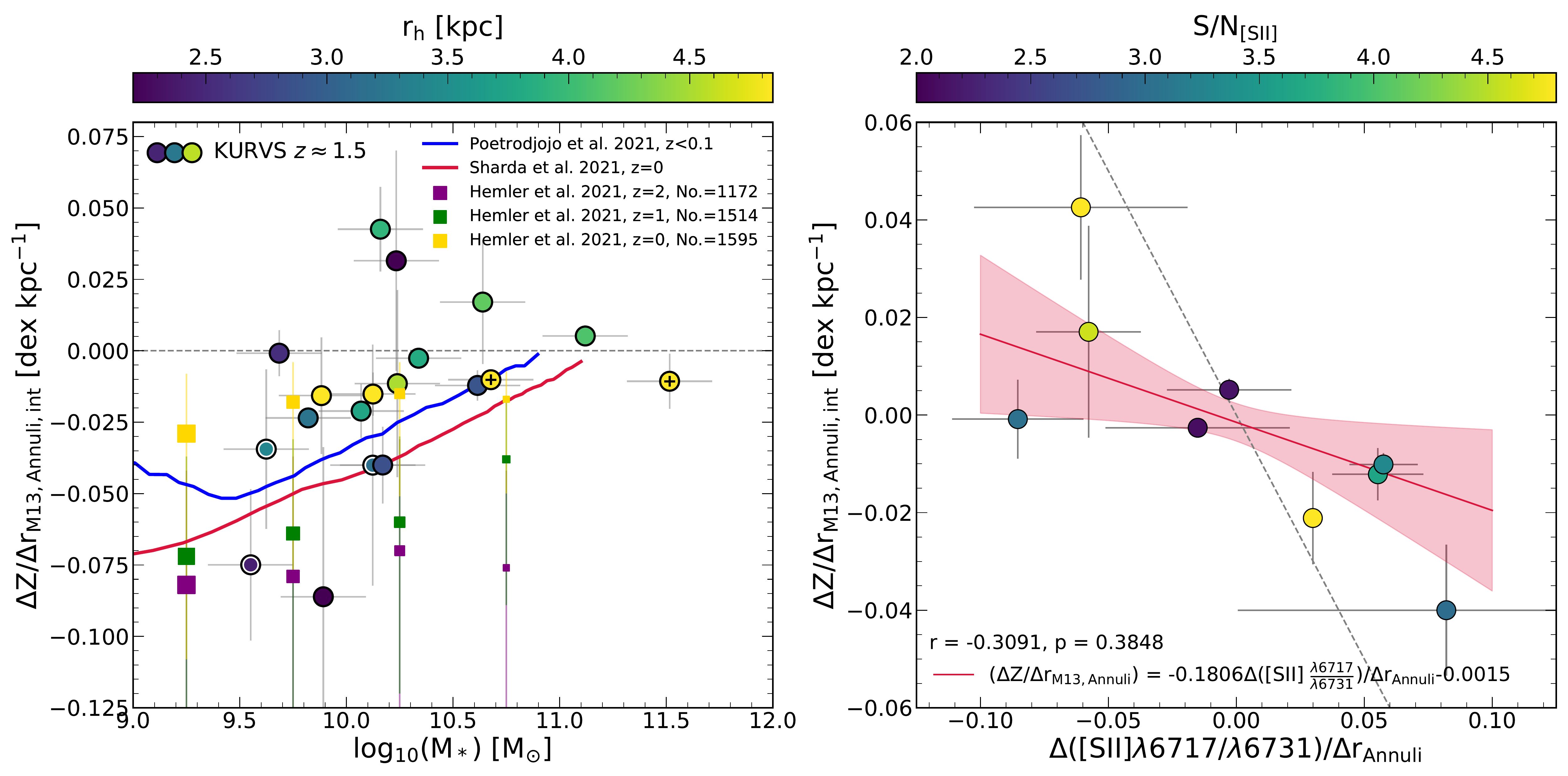}
    \caption{\textbf{Left:} The beam-smear corrected annuli-based metallicity gradient as a function of the galaxies stellar mass. We also show comparison samples from the SAMI survey at $z<0.1$ \citep{Poetrodjojo2021}, the analytical model of \citet{Sharda2021} and results from the TNG hydrodynamical simulations at redshift snapshots $z=1, 2, 3$ \citep{Hemler2021}. The size of the \citet{Hemler2021} data points reflect the number of galaxies in each stellar mass bin. The KURVS-CDFS galaxies have similar gas-phase metallicity gradients for given stellar mass as the low-redshift comparison samples becoming flatter at higher masses.  
    \textbf{Right:} The intrinsic annuli-based metallicity gradient as a function of the radial dependence of \SIIR{} ratio, coloured by the galaxy integrated \SII{} S/N. A spearman rank coefficient ($r_{\rm s}, p_{\rm s}$) and an orthogonal distance regression fit with 1$-\sigma$ uncertainty (crimson line and shaded region) are shown. The black dashed line indicates a one to inverse one relation . We identify an anti-correlation whereby a galaxy with a positive gas-phase metallicity  has a negative gradient in \SIIR{} (electron density).}
    \label{fig:OH_profiles}
\end{figure*}

In this way large numbers of metallicity gradient measurements can be made in distant galaxies with lower S/N spectra. \citep[e.g.][]{Wuyts2016,Schreiber2018,Gillman2021}. 
For the KURVS-CDFS sample we measure the annuli-based metallicity gradients utilising the \MT{} strong-line calibration (Section \ref{Sec:metals}) \col{ and the methods of \citet{Gillman2021}. In short we sum the spectra in the velocity-subtracted data cubes from elliptical annuli whose semimajor axes are multiplies of the half-light radius of the KMOS PSF. If three measurements of the \NH{} ratio can be made with a \HA{} S/N$>$3 in each annulus, the metallicity gradient is measured.} We derive a median metallicity gradient of \med{\zgrad}{$-$0.0052}{0.0018} dex\,kpc$^{-1}$. The PSF of the KMOS observations acts to flatten the observed metallicity gradients. To correct for this and attain an intrinsic metallicity gradient for each galaxy, we apply a beam-smearing correction to the gradients. The beam-smearing correction is a function of galaxy size (\Rh), relative to the FWHM of the observations, and galaxy axis ratio and was derived from modelling of mock galaxies with varying intrinsic gradients, continuum sizes and PSFs  \citep[see][]{Gillman_thesis,Gillman2021}. For the KURVS-CDFS sample we derive a median beam-smearing correction of \med{$C_{\rm BS}$}{0.32}{0.05} and median intrinsic metallicity gradient of \med{\zgrad}{$-$0.015}{0.005} dex\,kpc$^{-1}$. 

Recent studies have indicated connections between a galaxies gas-phase metallicity gradient and its fundamental properties such as stellar mass and  stellar continuum half-light radius \citep[e.g.][]{Hemler2021,Sharda2021,Sharda2021b,Sharda2021c,Boardman2021,Franchetto2021}. To analyse this correlation in the KURVS-CDFS sample, in Figure \ref{fig:OH_profiles} we correlate the intrinsic annuli-based  metallicity gradients with the galaxies stellar mass. We also show comparison samples from the SAMI survey at $z<0.1$ \citep{Poetrodjojo2021}, the analytical model of \citet{Sharda2021} and results from the TNG hydrodynamical simulations at redshift snapshots $z=0,1,2$ \citep{Hemler2021}. The intrinsic metallicity gradients of the KURVS-CDFS galaxies exhibit a similar correlation to the comparison samples, with flatter (or inverted) metallicity gradients at higher stellar mass. \citet{Sharda2021} predict that the up turn to flatter gradients is driven by the galaxies transitioning from an advection-dominated evolution (i.e re-distribution of metals) to a accretion-dominated evolution at higher stellar masses.

\citet{Boardman2021} identify a variation in the metallicity gradient of galaxies across the stellar mass - stellar continuum half-light radii plane. They establish that smaller galaxies have flatter gradients than larger galaxies for a given stellar mass. In Figure \ref{fig:OH_profiles} we colour the galaxies by their stellar continuum half-light radius (\Rh) however in the KURVS-CDFS sample no clear trend with galaxy size is seen.

\begin{figure*}
    \centering
    \includegraphics[width=\linewidth]{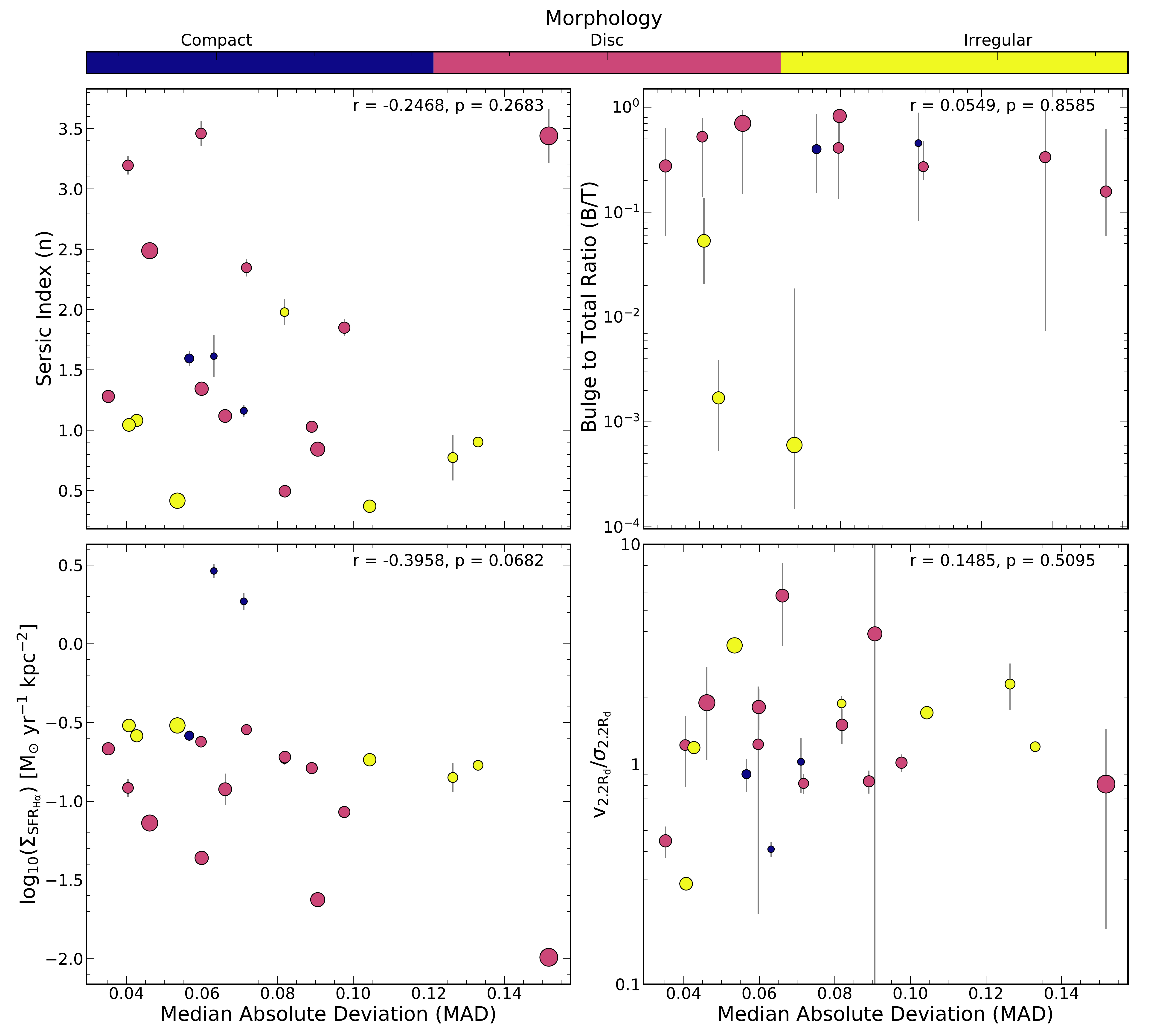}
    \caption{Median absolute deviation of the residuals to the linear pixel-based abundance profiles as a function of S\'ersic index, bulge to total ratio (B/T), H$\alpha$ star-formation rate surface density and rotation dominance (\vsig). The metallicity gradients are derived using the \citet{Marino2013} \NH{} calibration. We colour the galaxies by their visual rest-frame optical morphology following \citet{Huertas-Company2015} and the size of the data points reflect the F160W stellar continuum size derived by \citet{VanderWel2014}. In each panel we report the spearman rank ($r_{\rm s}, p_{\rm s}$) values, identifying only a negative correlation with star-formation surface density, whereby the metallicity profiles of galaxies with higher star-formation rate surface density have less deviation from  a linear profile.}
    \label{fig:grad_residual}
\end{figure*}

In addition to the radial variations in the gas-phase metallicity, other interstellar medium have been identified in local galaxies to correlate with radius. Studies have shown that galaxies have clumps or gradients in electron density \citep[e.g.][]{Binette2002, Phillips2007,Herrera-Camus2016} which brings into question the assumption of fixed electron density when inferring the properties of the interstellar medium (e.g. metallicity). To analyse the spatial variation in electron density in the KURVS-CDFS sample, we measure the \SIIR{} in annuli in order to increase the \SII{} S/N. Using the de-redshifted KMOS data cubes we apply the same method as for measuring the annuli-based metallicity gradients but this time measure \SIIR{} in each annulus. We require S/N$>$2 for the \col{weaker} \SII{} in each annuli and derive a median intrinsic \SIIR{} gradient for the 11 galaxies that fulfil this criteria of \med{$\Delta $(\SIIR)/$\Delta r$}{$-$0.007}{0.034} dex\,kpc$^{-1}$.

In Figure \ref{fig:OH_profiles} we correlate the intrinsic \SIIR{} gradients with the intrinsic annuli-derived metallicity gradients using the \MT{} calibration. We perform an orthogonal distance regression fit (crimson line) of the form \zgrad\,=\,$\alpha$\,(\SIIR)/$\Delta r$+\,$\beta$ deriving \med{$\alpha$}{$-$0.18}{0.08} and \med{$\beta$}{$-$0.0015}{0.004}. A negative correlation with a spearman rank coefficient of $r_{\rm s}$\,=\,$-$0.31, $p_{\rm s}$\,=\,0.39 is observed, suggesting an anti-correlation between the electron density and metallicity gradient. In this instance galaxies with negative abundance gradients have positive gradients in electron density. \col{This brings into question the use of strong-line calibrations that assume a fixed electron density in the interstellar medium. Variations in electron density maybe caused by variations in star-formation rate, midplane and  feedback pressures of the interstellar medium as well as variations in molecular hydrogen density \citep[e.g.][]{Kakkad2018,Davies2020,Davies2021}.} However larger samples are required to confirm this correlation at higher significance. 

\subsubsection{Evidence for non-uniform gradients}

Main sequence high-redshift galaxies have been observed to have elevated velocity dispersions \citep[e.g.][]{Schreiber2006,Wisnioski2015,Johnson2018} and non-uniform clumpy rest-frame optical morphologies \citep[e.g.][]{Glazebrook1995,Abraham1999,Conselice2014,Harrison2017}. A linear metallicity gradient therefore appears as an over simplification of a much more complex abundance profile within these galaxies (see Figure \ref{fig:HST_resolved}).  

Several studies have shown evidence for azimuthal variations in a galaxies gas-phase metallicity both at low redshift \citep[e.g.][]{Sanchez_meng2018,Sanchez2021,Li2021,Metha2021}  and at high-redshift \citep[e.g.][]{Schreiber2018,Curti2020}  with connections to the galaxies morphology and kinematic state. To analyse the non-linearity of the abundance profiles in high-redshift star-forming galaxies we use the spatially resolved \NH{} emission-line maps derived in \AP{} (Figure \ref{fig:HST_resolved}). Correlating the metallicity of each pixel with its de-projected radius, we define a pixel-based metallicity gradient for each galaxy, deriving a median gradient of of \med{\zgrad}{$-$0.0025}{0.0010} dex\,kpc$^{-1}$ for the sample. 

For each galaxy we subtract off the linear pixel-based metallicity gradient and define the median absolute deviation (MAD) of the residuals; where a larger MAD indicates a larger derivation from the linear abundance profile. The median MAD from a linear abundance profile for KURVS-CDFS sample is \med{MAD}{0.07}{0.01}\,dex with a 16th\,--\,84th percentile range of 0.04\,--\,0.10\,dex. To analyse the correlation between the metallicity deviations from linearity and the galaxies properties we use the KMOS data to quantify the galaxies kinematic properties as well as the CANDELS \textit{HST} imaging to constrain the morphological properties of the galaxies. 

The shape of the stellar continuum light profile was quantified by \citet{VanderWel2014} who used the 1.6$\mu$m \textit{HST} imaging to constrain the S\'ersic index of the CANDELS galaxies. Fitting S\'ersic profiles with the \galfit{} software to 10 arcsecond cutouts of the galaxies they established a median S\'ersic index of \med{$n$}{1.23}{0.21} for the KURVS-CDFS galaxies. In addition we have use the bulge to total (B/T) ratios for the KURVS-CDFS galaxies from \citet{Dimauro2018}, who used the multi-wavelength \textit{HST} data to decompose each galaxy into its bulge and disc components. From this sample, the KURVS-CDFS galaxies have a median bulge to total ratio of \med{B/T}{0.27}{0.12} i.e predominantly disc-dominated.

A visual classification of the galaxies was carried out by \citet{Huertas-Company2015}, using machine learning to distinguish between Discs, Compact and Irregular galaxies. Following this classification the KURVS-CDFS sample is composed of 12 Discs, 9 Irregular and 3 Compact galaxies. In Figure \ref{fig:grad_residual} we correlate the MAD of the metallicity profiles with the  S\'ersic index, B/T, colouring the galaxies by their Hubble morphology (Compact, Disc, Irregular).

We also correlate the median absolute deviation of the residuals to the linear abundance profiles with the galaxies \HA{} star-formation rate surface density ($\Sigma_{\rm SFR}$) and  kinematic state (\vsig) derived from the KMOS observations (see \AP{} for details). In Figure \ref{fig:grad_residual} we only identify a negative correlation between the deviation from a linear metallicity profile and the $\Sigma_{\rm SFR}$ with spearman rank coefficients of $r_{\rm s}$\,=\,$-$0.40, $p_{\rm s}$\,=\,0.07. 

This implies  that although there are clear azimuthal variations in the metallicity of the interstellar medium in high-redshift galaxies (Figure \ref{fig:HST_resolved}), they are not necessarily connected to the galaxies rest-frame optical morphology or kinematic properties of the galaxies but are more related to the concentration of ongoing star formation. At a fixed galaxy integrated SFR$_{\rm H\alpha}$, larger galaxies have more deviation from a linear abundance gradient, whilst for a fixed size, galaxies with lower star-formation rates also have a larger deviation from a linear gradient. To visualise this in Figure \ref{fig:grad_residual}, the size of the data points reflect the stellar continuum size of the galaxies. A similar connection between azimuthal variations in gas-phase metallicity and main-sequence properties was identified by \citet{Li2021} in the local Universe, with no correlation with galaxies kinematics or morphology. \col{To investigate further, we derive the residuals in metallicity, stellar mass surface density and star formation rate surface density to their respective linear profiles as a function of radius for the KURVS-CDFS sample. We identify no correlation between the residuals of the metallicity profiles and the stellar mass surface density and star formation rate surface density residuals with $r\,=\,0.0360,p\,=\,0.1137$, and  $r\,=\,-0.0267,p\,=\,0.2409$ respectively.}

\col{Defining the MAD at each radius we can construct the MAD radial profile for metallicity, stellar mass surface density and star formation rate surface density residuals. All three MAD profiles show an increase with radius (see Appendix \ref{App:MAD}), indicating larger deviations from linear profiles at larger radii. However, the variations in the MAD of the residuals to the stellar mass surface density and star formation rate surface density profiles are not reflected in the MAD of the metallicity profile. This indicates that deviations from a linear metallicity gradient are not correlated with deviations from a linear stellar mass surface density or star formation rate surface density profile, at the same radius.}

\section{Conclusions}\label{Sec:Conc}

In this paper we have analysed the interstellar medium properties of 22 star-forming galaxies from the first half of the KMOS Ultra-deep Rotational Velocity Survey (KURVS) which comprises of sources in the CDFS field (KURVS-CDFS). KURVS-CDFS galaxies have a medium redshift of \zest{1.39}. We demonstrate the galaxies are representative of the galaxy main-sequence at this epoch with a median stellar mass, \HA{} derived star-formation rate and stellar continuum half-light radius of \med{$\log_{10}$(\Mstar[$M_{\odot}$])}{10.10}{0.09}, \med{SFR$_{\rm H\alpha}$[M$_{\odot}$yr$^{-1}$]}{24}{6} and, \med{\Rh[kpc]}{3.28}{0.42} respectively. Our main conclusions are;
\begin{itemize}
    \item The stellar mass\,--\,gas-phase metallicity relation is in place in the sample with the galaxies exhibiting integrated gas-phase metallicity values expected for this epoch with a median value of \med{12+log(O/H)$_{\rm M13}$}{8.40}{0.03}. We establish galaxies with higher \SH{} ratio show closer agreement between the \citet{Dopita2016} \NS{} based metallicity calibration and \citet{Marino2013} \NH{} calibration, especially at high metallicity (Figure \ref{fig:integrated_metals}). Whilst at low metallicity the integrated \SH{} ratio of the galaxies shows no correlation with the offset between the metallicity indicators.
    \item We quantify the electron density in the galaxies using the \SIIR{} ratio. Of the 12 galaxies with 0.45\,$\leq$\,\SIIR$\,\leq$\,1.45 we derive a median value of \med{log$_{10}$(\nel[cm$^{-3}$])}{2.29}{0.25}. We demonstrate the electron density within the sample is elevated in comparison to local galaxies in agreement with other high-redshift studies. However, we find no correlation with the ionization parameter (as traced by the integrated \SH{} ratio) (Figure \ref{fig:ne_evolution}) as well as no correlation between the galaxies main-sequence properties (e.g stellar mass, star-formation rate, specific star-formation rate) and electron density (Figure \ref{fig:ne_prop}).
    \item Exploiting the high S/N spatially resolved KMOS and \textit{HST} data (Figure \ref{fig:HST_resolved}) we define the resolved fundamental metallicity relation (rFMR)  at \zest{1.5} identifying the correlation between gas-phase metallicity, stellar mass surface density and \HA{} star-formation rate surface density (Figure \ref{Figure:resolvedMZR}). We find a 0.05\,$\pm$\,0.01\,dex evolution in the normalisation of the relation compared to the local Universe  as well as identifying variation in the gas-phase metallicity as a function of radial position.
    \item We define stellar mass gas-phase metallicity gradient relation in the KURVS-CDFS sample, identifying flatter metallicity gradient at higher stellar masses as found by low redshift studies (Figure \ref{fig:OH_profiles}). We further define the relation between the \SIIR{} ratio (electron density) gradient and gas-phase metallicity gradient in the galaxies establishing a  spearman rank coefficient of $r_{\rm s}$\,=\,$-$0.31, $p_{\rm s}$\,=\,0.39, suggesting an anti-correlation between the electron density and metallicity gradient. In this instance galaxies with negative abundance gradients have positive gradients in electron density, however larger samples are required to confirmed this at higher significance. 
    \item Finally we examine the non-linearity of the abundance profiles using the spatially-resolved \NH{} maps, identifying evidence for azimuthal variations in metallicity which show a negative correlation with galaxy integrated star-formation rate surface density ($r_{\rm s}\,$\,=\,$-$0.40,\,$p_{\rm s}$\,=\,0.07). However, no correlation is identified with the galaxies morphological and kinematic properties (Figure \ref{fig:grad_residual}) \col{We further identify no connection between deviations from a linear metallicity profile and the deviations from a linear stellar mass surface density or star formation rate surface density profile (see Appendix \ref{App:MAD}). Thus indicating that radial variations in metallicity are not connected to the radial variations in stellar mass or star formation but correlate with the global density of star-formation as traced by the \HA{} emission-line. }
\end{itemize}

Overall we have shown that \col{KURVS-CDFS sample at \zest{1.5} is composed of main-sequence galaxies which follow} the galaxy-integrated scaling relations observed at low redshift, between stellar mass, stellar continuum size, star-formation rate and gas-phase metallicity. \col{Whilst} their spatially-resolved interstellar medium properties indicate both radial and azimuthal variations. Consequently, the prescriptions used to define the fundamental properties of low-redshift galaxies may not always be applicable at high-redshift due to variations in interstellar medium properties (e.g. electron density, ionisation parameter and variations in the ionising source) which can be quantified by future observations with \textit{JWST}.

\FloatBarrier
\section*{Acknowledgements}

We thank D. Kashino and J.K. Barrera-Ballesteros for the availability of comparison sample emission-line ratios via private communication. We also thank the referee for a constructive review.
SG acknowledges the support of the Cosmic Dawn Center of Excellence funded by the Danish National Research Foundation under then grant 140. 
AMS and AP gratefully acknowledges financial support from STFC through grants ST/T000244/1 and ST/P000541/1.
UD acknowledges the support of STFC studentship ST/R504725/1.
This work was supported by the National Science Foundation of China (11721303, 11991052) and the National Key R\&D Program of China (2016YFA0400702).
This research made use of Astropy\footnote{http://www.astropy.org} a community-developed core Python package for Astronomy \citep{Astropy2013,Astropy2018}.

%%%%%%%%%%%%%%%%%%%%%%%%%%%%%%%%%%%%%%%%%%%%%%%%%%
\section*{Data Availability}

The inclusion of a Data Availability Statement is a requirement for articles published in MNRAS. Data Availability Statements provide a standardised format for readers to understand the availability of data underlying the research results described in the article. The statement may refer to original data generated in the course of the study or to third-party data analysed in the article. The statement should describe and provide means of access, where possible, by linking to the data or providing the required accession numbers for the relevant databases or DOIs.



\bibliographystyle{mnras}
\bibliography{master} % if your bibtex file is called example.bib

%\begin{thebibliography}{99}
%\end{thebibliography}

%%%%%%%%%%%%%%%%%%%%%%%%%%%%%%%%%%%%%%%%%%%%%%%%%%

%%%%%%%%%%%%%%%%% APPENDICES %%%%%%%%%%%%%%%%%%%%%

\appendix

\begin{figure*}
\section{Spatially Resolved Properties}\label{App:Resolved_Properties}
\centering
    \includegraphics[width=\linewidth]{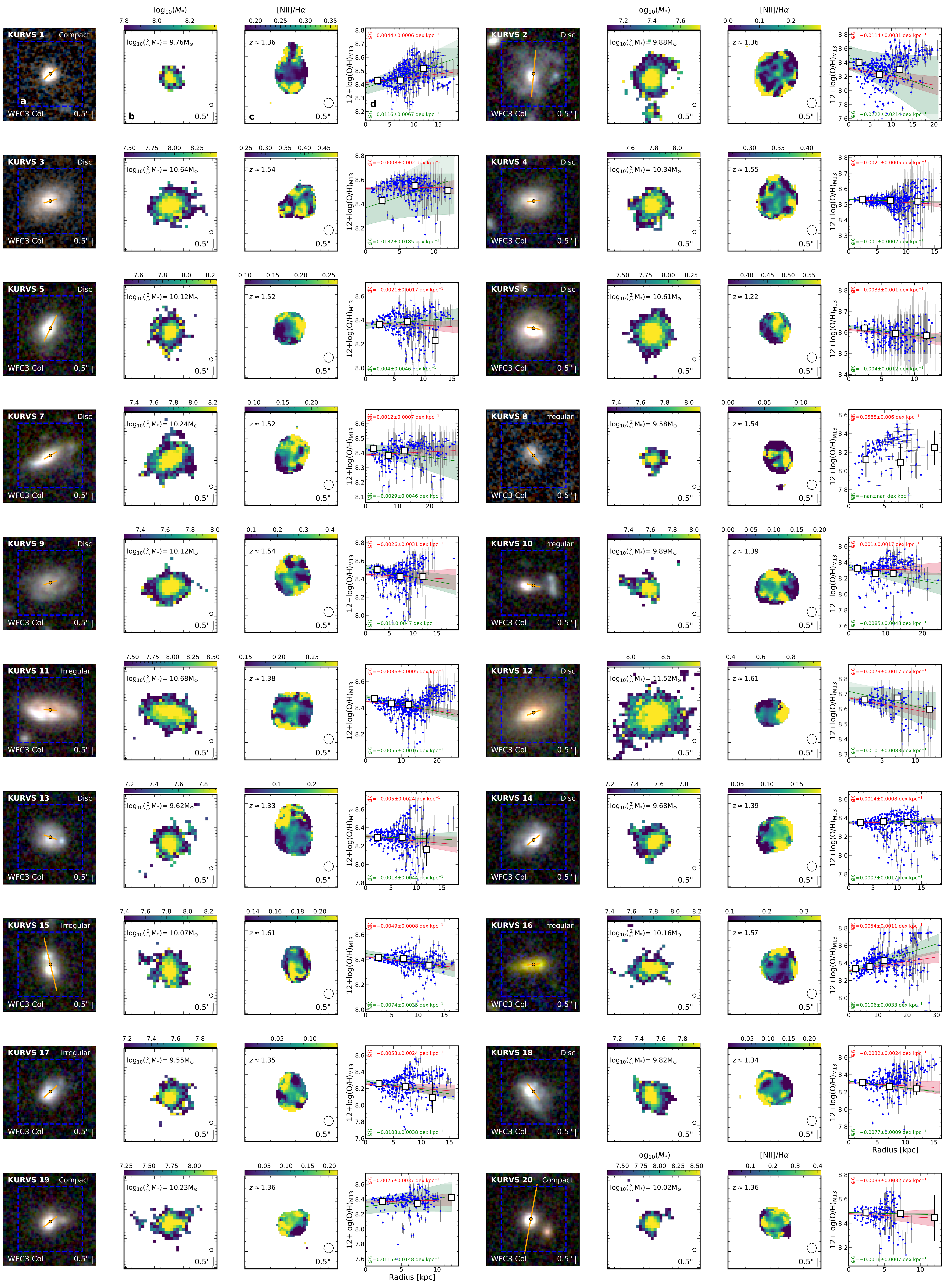}
    \label{fig:mass_map_P1}
\end{figure*}

\begin{figure*}
\centering
    \includegraphics[width=\linewidth, trim = {0cm 12cm 0cm 0cm}]{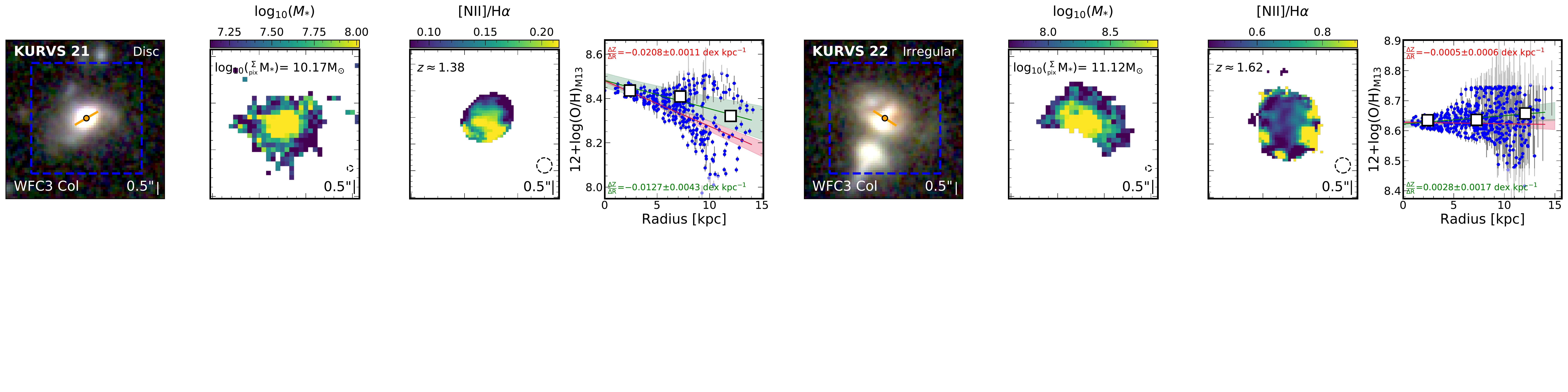}
    \caption{Examples of the \textit{HST} multi-band imaging and spatially resolved KMOS data for the KURVS-CDFS galaxies. For each galaxy we show a WFC3 three-colour image composed of F105W, F125W and F160W images with the KMOS field of view (blue square) overlaid (\textbf{a}). The semi-major axis of the galaxy (orange line) and stellar continuum centre (black filled circle) are indicated, as well as the morphological classification from \citet{Huertas-Company2015}. The spatially resolved stellar mass map  (\textbf{b}) and \NH{} map  (\textbf{c}) are shown with the FWHM of the PSF in \textit{HST} and KMOS  observations (black dashed circle) respectively. In the final panel (\textbf{d}), we display the pixel-based gas-phase metallicity (blue points) as a function of radius and linear fit (crimson line) with reported slope and uncertainty. We also show the annuli-based metallicity (black squares) and corresponding annuli-based gradient (green line) with slope and uncertainty. The examples indicate the range of morphologies and metallicity profiles in addition to the KMOS data quality.}
    \label{fig:mass_map_P2}
\end{figure*}

\begin{figure*}
\section{Median Absolute Deviation (MAD)
Profiles}\label{App:MAD}
\begin{flushleft}
To further investigate the deviations from a linear abundance profile in the KURVS-CDFS galaxies, we define the radial profiles in each galaxy for metallicity, stellar mass surface density and star formation rate surface density. Performing a linear profile to each profile, for each galaxy, we define the residuals to each profile. Taking the sample as we whole, we then define the MAD of the residuals as a function of radius. 
\newline

In Figure  \ref{fig:MAD} we show the MAD as function of radius for metallicity, stellar mass surface density and star formation rate surface density. All three MAD profiles show an increase with radius, indicating larger deviations from linear profiles at larger radii. However, the variations in the MAD of the residuals to the stellar mass surface density and star formation rate surface density profiles are not reflected in the MAD of the metallicity profile. This indicates that deviations from a linear metallicity gradient are not correlated with deviations from a linear stellar mass surface density or star formation rate surface density profile, at the same radius.
\end{flushleft}
\centering
    \includegraphics[width=0.7\linewidth]{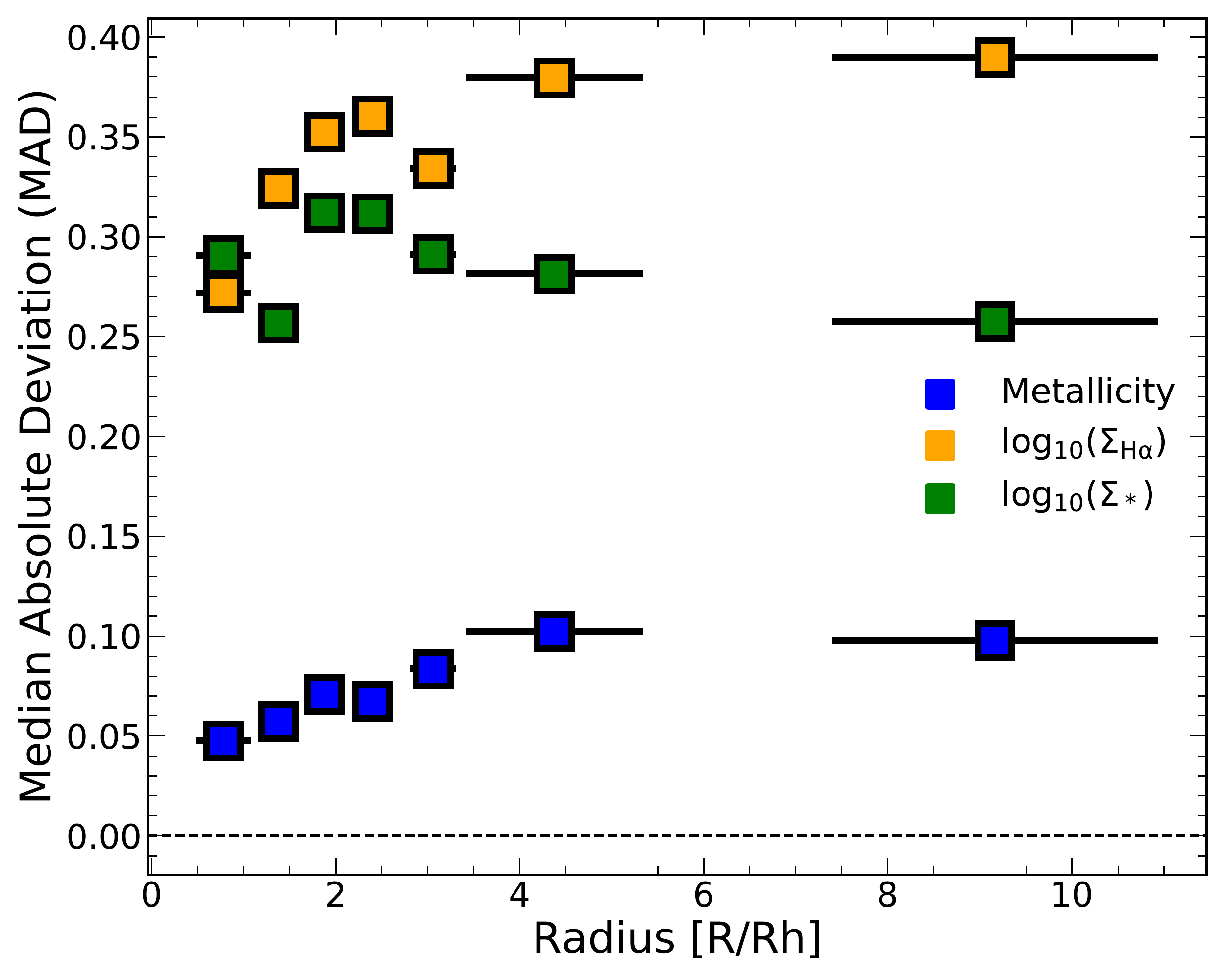}
    \caption{The median absolute deviation (MAD) of the residuals as a function of radius for the metallicity, stellar mass surface density and star formation rate surface density profiles in the KURVS-CDFS sample. Each radial bins contains 300 pixels. All three profiles show an increase in MAD with radius. The variations in stellar mass surface density residual MAD are reflected in the star formation rate surface MAD but not in the metallicity MAD profile. This indicates that deviations from a linear metallicity gradient are not correlated with deviations from a linear stellar mass surface density or star formation rate surface density profile, at the same radius.}
    \label{fig:MAD}
\end{figure*}

%%%%%%%%%%%%%%%%%%%%%%%%%%%%%%%%%%%%%%%%%%%%%%%%%%

% Don't change these lines
\bsp	% typesetting comment
\label{lastpage}
\end{document}